\documentclass[sigconf, nonacm]{acmart}

\title[Galerkin Method of Regularized Stokeslets for Procedural Fluid Flow with Control Curves]{Galerkin Method of Regularized Stokeslets\\for Procedural Fluid Flow with Control Curves}
\author{Ryusuke Sugimoto}
\orcid{0000-0001-5894-0423}
\affiliation{
    \institution{University of Waterloo}
    \city{Waterloo}
    \state{Ontario}
    \country{Canada}
}
\email{rsugimot@uwaterloo.ca}

\author{Jeff Lait}
\orcid{0009-0004-7114-9168}
\affiliation{
    \institution{Side Effects Software Inc.}
    \city{Toronto}
    \state{Ontario}
    \country{Canada}
}
\email{jlait@sidefx.com}

\author{Christopher Batty}
\orcid{0000-0003-3830-7772}
\affiliation{
    \institution{University of Waterloo}
    \city{Waterloo}
    \state{Ontario}
    \country{Canada}
}
\email{c2batty@uwaterloo.ca}

\author{Toshiya Hachisuka}
\orcid{0000-0003-0284-3776}
\affiliation{
    \institution{University of Waterloo}
    \city{Waterloo}
    \state{Ontario}
    \country{Canada}
}
\email{thachisu@uwaterloo.ca}

\usepackage{cleveref}

\newcommand{\pres}{p}
\newcommand{\vel}{\mathbf{u}}

\newcommand{\force}{\mathbf{f}}
\newcommand{\torque}{\tau}
\newcommand{\bodyf}{\mathbf{b}}
\newcommand{\curve}{\mathcal{C}}
\newcommand{\vecx}{\mathbf{x}}
\newcommand{\vecy}{\mathbf{y}}
\newcommand{\vecr}{\mathbf{r}}
\newcommand{\vect}{\mathbf{t}}
\newcommand{\dsx}{\,\mathrm{d}s_\vecx}
\newcommand{\dsy}{\,\mathrm{d}s_\vecy}
\newcommand{\re}{r_\epsilon}

\newcommand{\Rtwo}{\mathbb{R}^2}
\newcommand{\Rthree}{\mathbb{R}^3}
\newcommand{\identity}{\mathbf{I}}
\newcommand{\transpose}{^\mathsf{T}}
\newcommand{\stokeslet}{\mathbf{S}}
\newcommand{\regularized}{^\epsilon}

\crefformat{figure}{Fig.~#2#1#3}
\crefformat{equation}{Eq.~#2#1#3}

\begin{teaserfigure}
\centering
\includegraphics[width=0.28\linewidth]{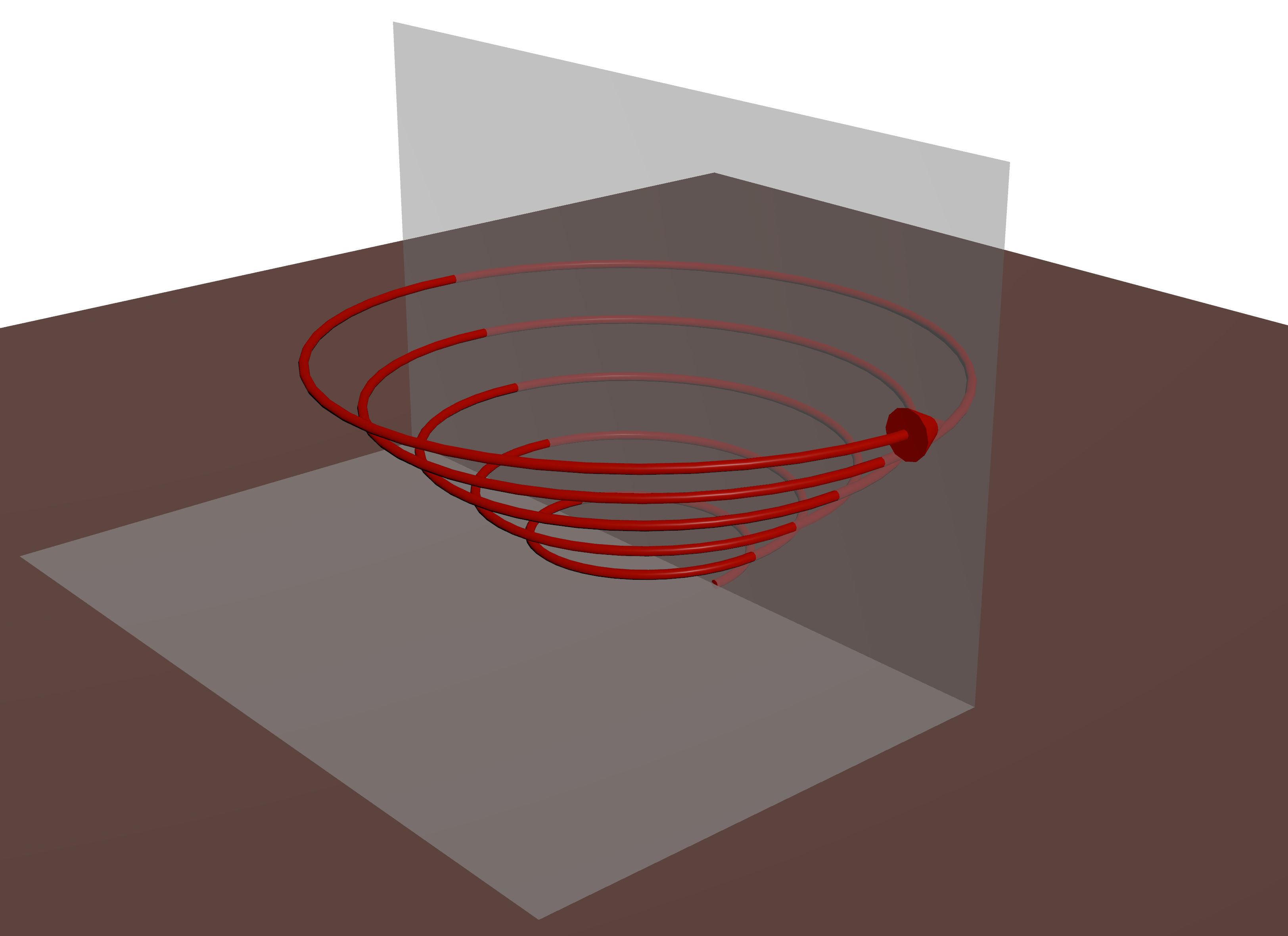}%
\includegraphics[width=0.28\linewidth]{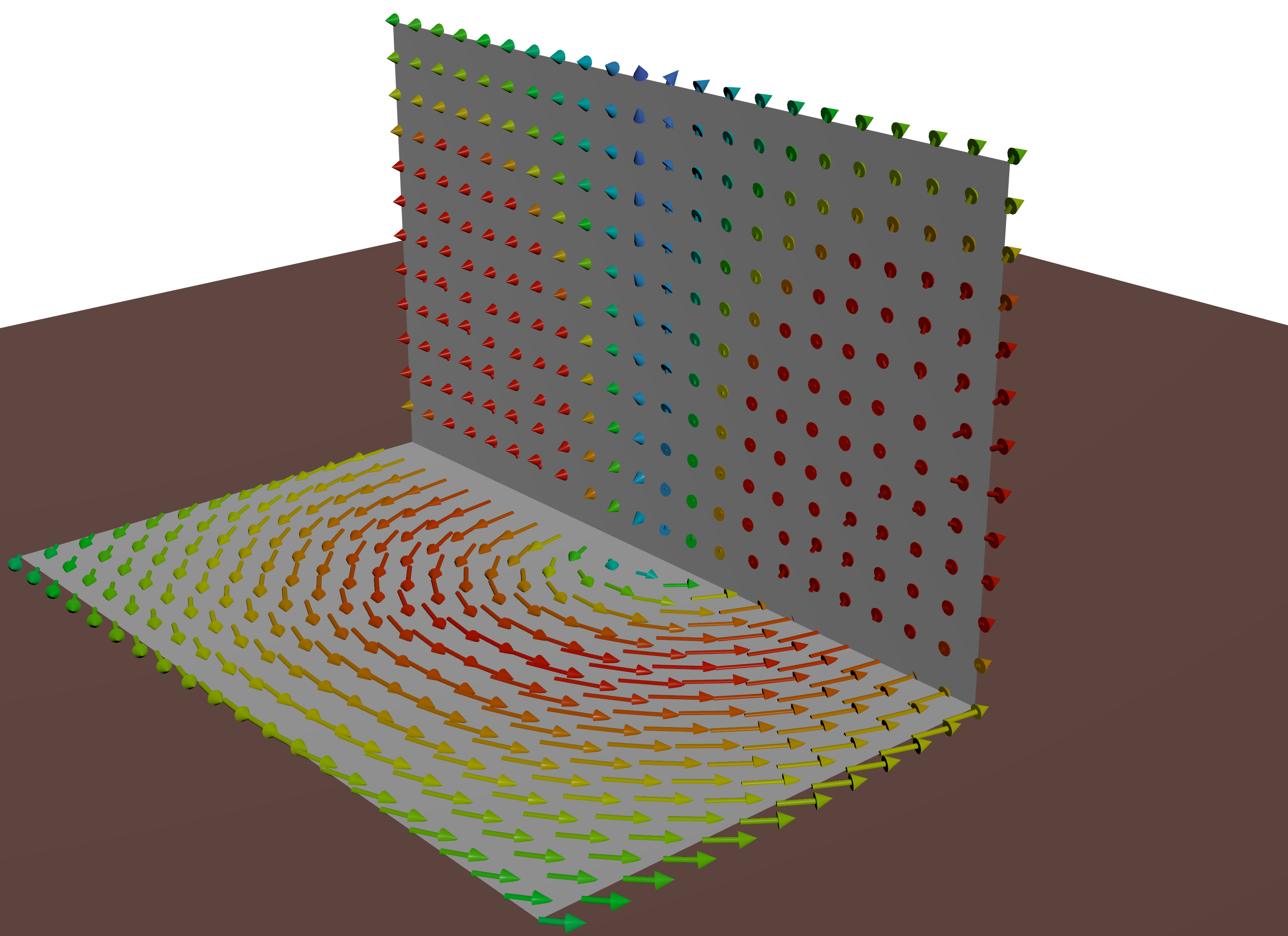}%
\begin{minipage}[b]{0.43\linewidth}
    \includegraphics[width=0.33\linewidth]{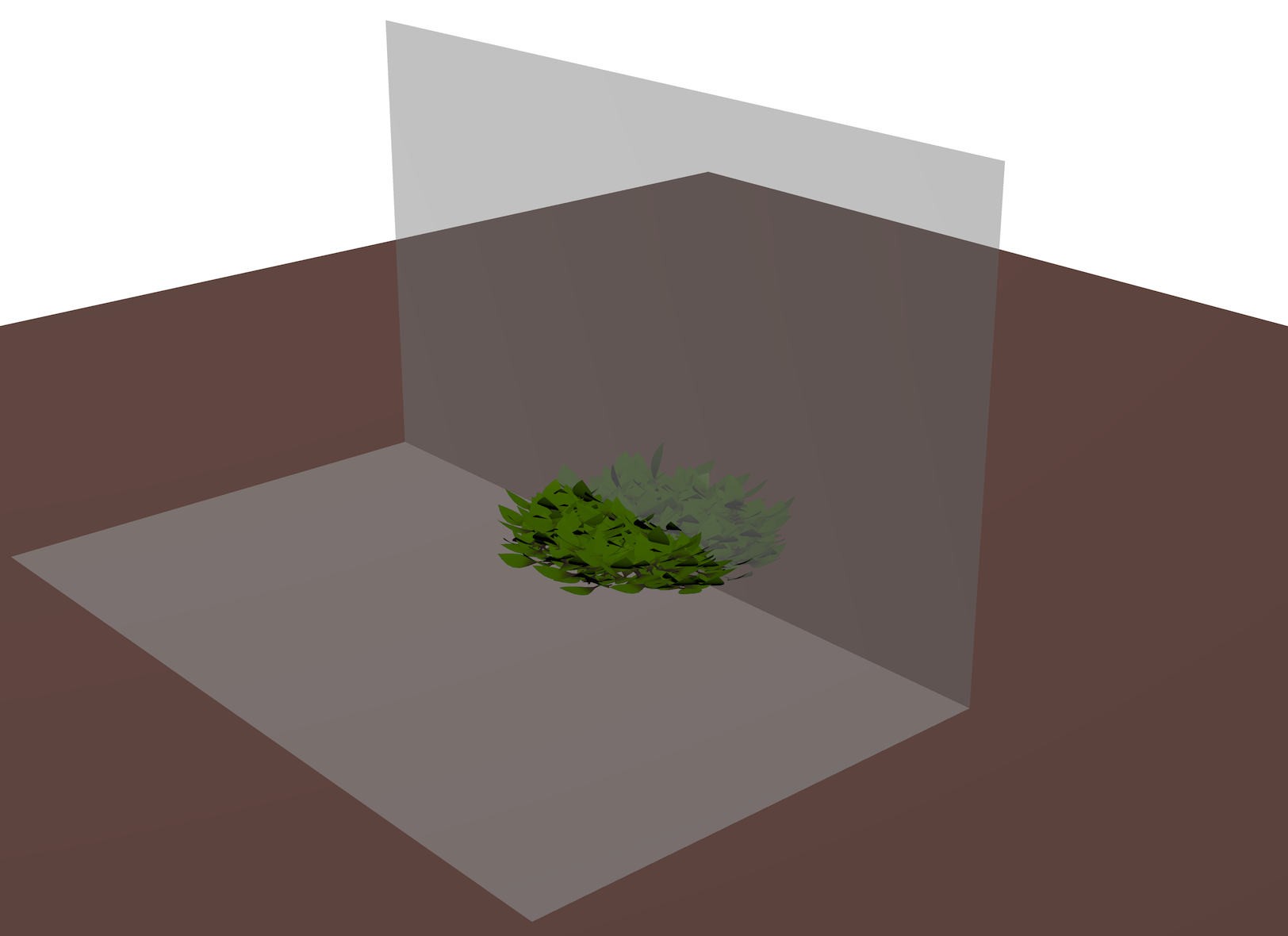}%
    \includegraphics[width=0.33\linewidth]{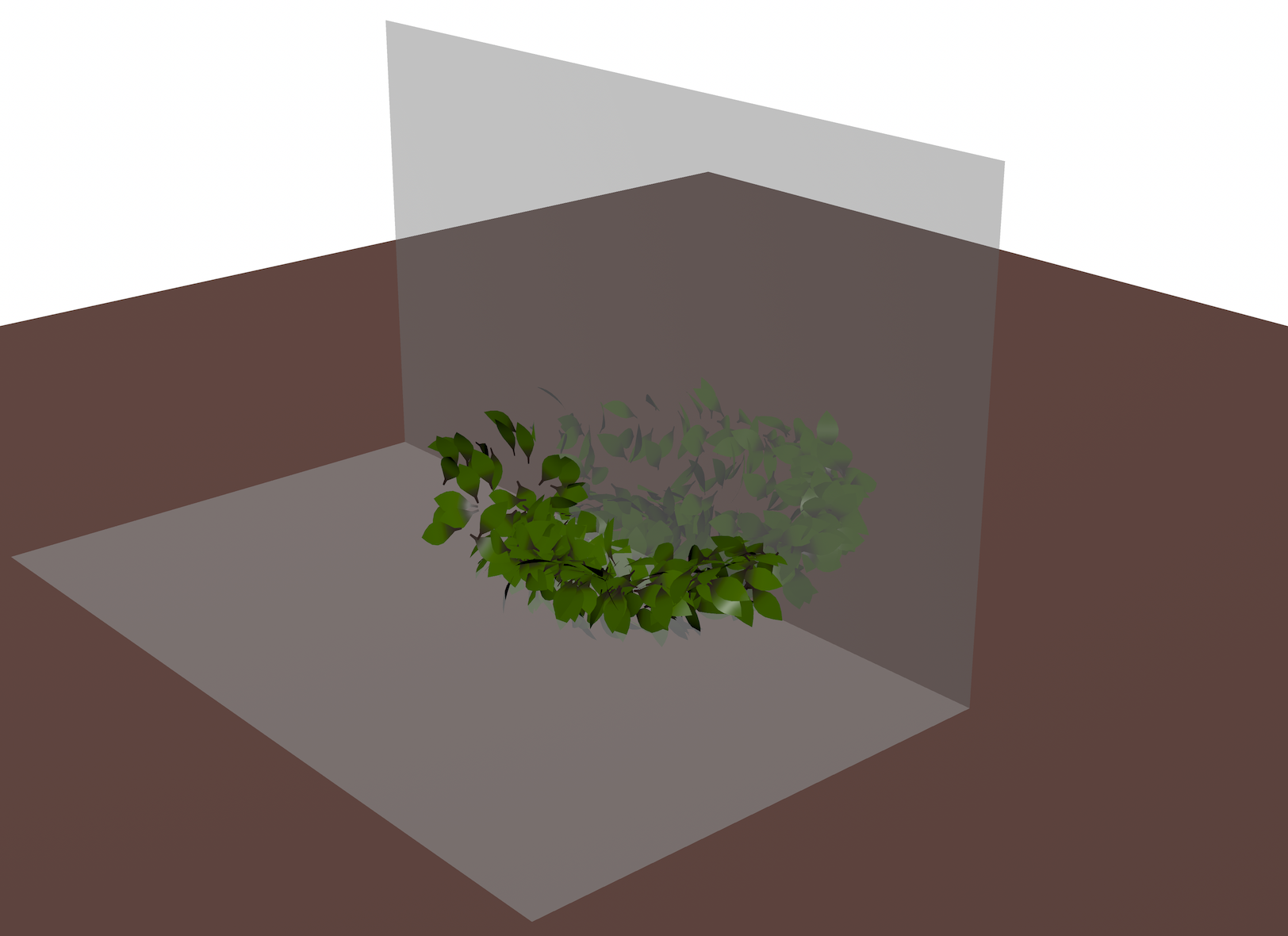}%
    \includegraphics[width=0.33\linewidth]{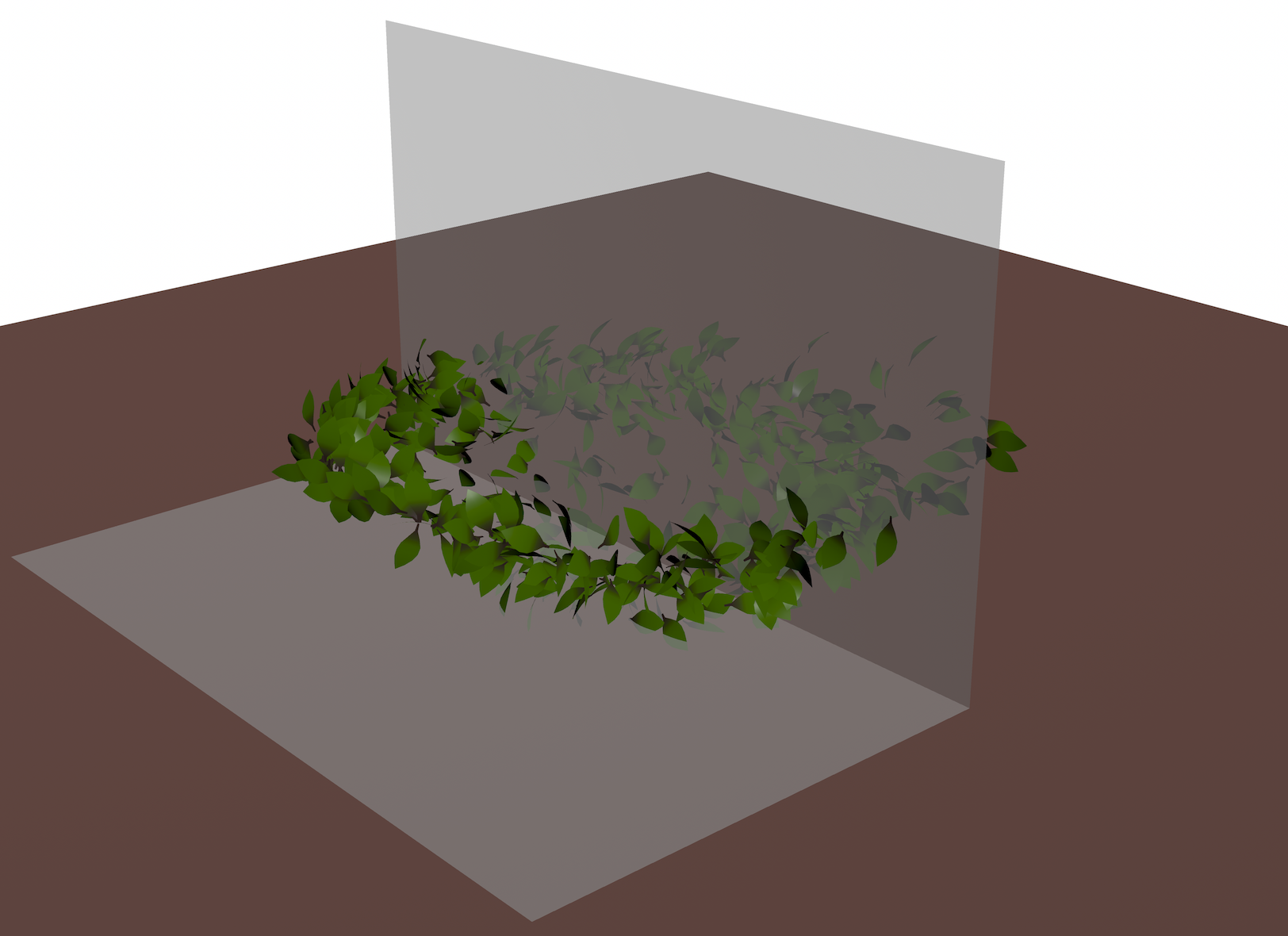}
    \includegraphics[width=0.33\linewidth]{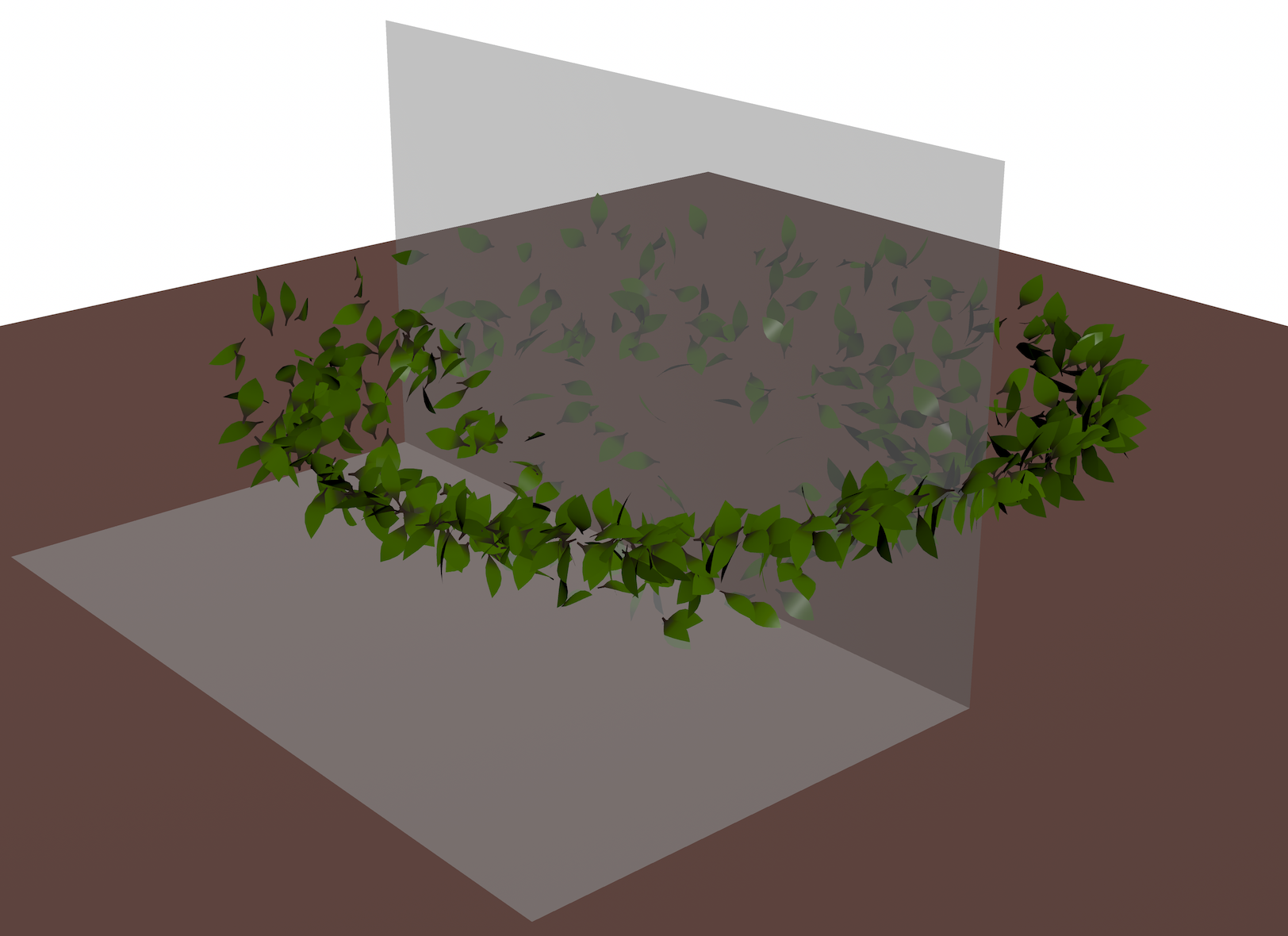}%
    \includegraphics[width=0.33\linewidth]{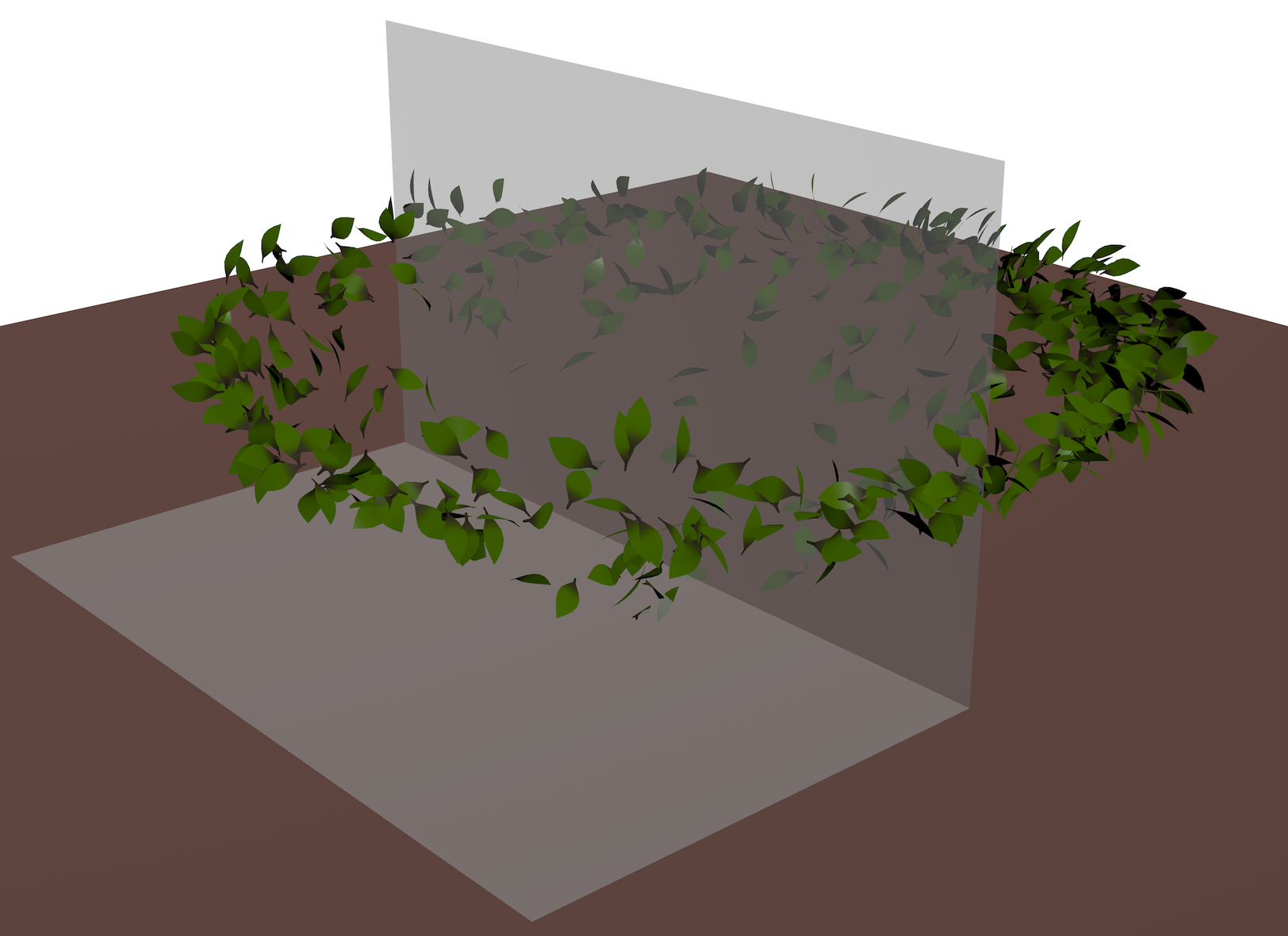}%
    \includegraphics[width=0.33\linewidth]{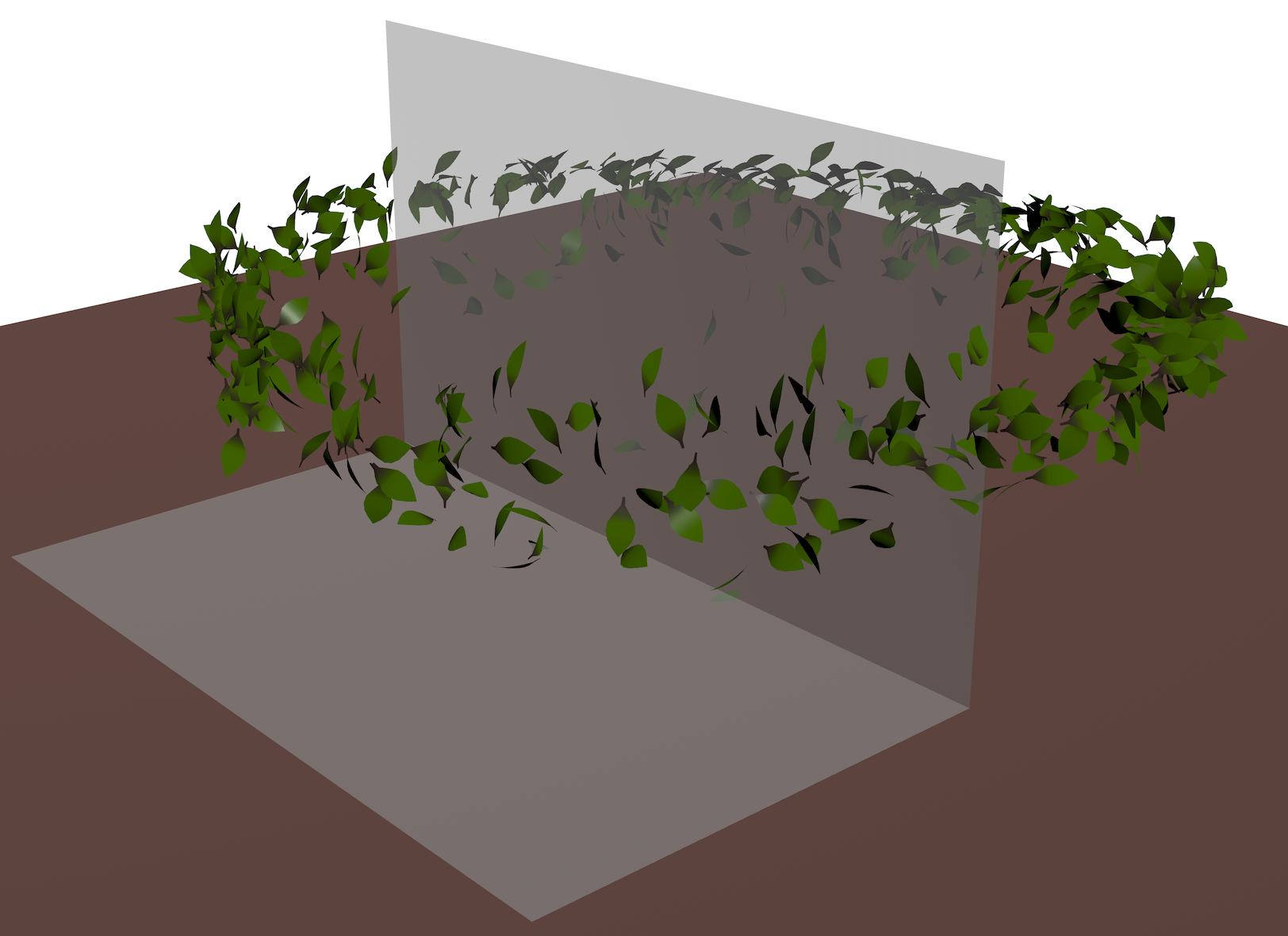}
\end{minipage}
\caption{Leaves caught up in a whirlwind. We manually specify the velocity and angular velocity along the control curve (left). The method can compute the resulting incompressible velocity field at any point in the ambient space, such that it conforms to the input along the control curve (middle). We evaluate the velocity at the leaf positions to advance the particle-based animation of leaves in a whirlwind (right). See the supplemental video for animation.}
\label{fig:teaser}
\end{teaserfigure}

\begin{abstract}
    We present a new procedural incompressible velocity field authoring tool, which lets users design a volumetric flow by directly specifying velocity along control curves. Our method combines the Method of Regularized Stokeslets with Galerkin discretization. Based on the highly viscous Stokes flow assumption, we find the force along a given set of curves that satisfies the velocity constraints along them. We can then evaluate the velocity anywhere inside the surrounding infinite 2D or 3D domain. We also show the extension of our method to control the angular velocity along control curves. Compared to a collocation discretization, our method is not very sensitive to the vertex sampling rate along control curves and only requires a small linear system solve.
\end{abstract}



\begin{document}
\maketitle

\section{Introduction}
Authoring a plausible incompressible fluid velocity field from user inputs is a longstanding, challenging problem.
Incompressible (i.e., divergence-free)  fields are strongly preferred because a divergent velocity implies the presence of nonphysical sources or sinks of fluid material. 
There are currently two popular options for authoring incompressible velocity fields: projection of an input velocity field and Curl-Noise~\cite{Bridson2007CurlNoise}.

The projection approach solves Poisson's equation to remove divergence in the input velocity field, thereby making it incompressible. This option is implemented in many fluid simulation systems and necessitates discretization. Discretization of a given velocity field can be tedious for artists to set up, and depending on the spatial resolution, the result can exhibit large errors compared to the analytical solution. Moreover, projection often alters the input velocity field in an undesirable and unpredictable manner, making it difficult for users to achieve their intended flow.

The Curl-Noise approach~\cite{Bridson2007CurlNoise} essentially constructs a velocity field as the curl of a user-designed potential function.
Unlike the first option, Curl-Noise does not involve any globally coupled linear system (e.g., Poisson's equation) and does not require discretization. While adding a turbulent, incompressible noise to an input velocity field is easy with Curl-Noise, it remains highly unintuitive for users to specify an intermediate potential field whose curl would yield a desired incompressible flow.

We propose a new velocity field design approach: we let users specify the velocity \emph{directly} along control curves, and then naturally extend those velocities throughout the ambient space. This approach contrasts with the indirect control offered by Curl-Noise, and, unlike the projection method, does \emph{not} discretize the volumetric ambient space.
Our method extends the Method of Regularized Stokeslets (MRS)~\cite{Cortez2001Stokeslets, Cortez2005Stokeslets3D} to achieve this.
We represent the resulting velocity field with a superposition of (regularized) \emph{Stokeslets}, which are \emph{fundamental solutions} for the steady-state \emph{Stokes equations}. The Stokes equations model the velocity field for highly viscous and incompressible quasi-static flows. A Stokeslet represents a solution to the equation in an infinite domain under a point-concentrated applied force. Given the velocities at a few sample points, we can solve a linear system to determine the forces at those points that satisfy the user's velocity constraints. Then, by superposing Stokelets at those locations with corresponding forces, we can compute a velocity field at any point inside the domain.
The standard MRS assumes that the velocity is specified at a few independent points. We extend MRS to allow the velocity to be specified along control curves represented as polylines and construct a linear system with a Galerkin method to intuitively control the resulting velocity field.  
We also adapted the twist control introduced for elasticity~\cite{DeGoes2017Kelvinlets} so that users can specify the angular velocity of the field along the control curves. A similar control is available in force field design tools~\cite{Houdini, Blender}, and having the capability to directly design the angular velocity improves the usability of our tool.
An implementation of our algorithm is shipped as a new feature (\href{https://www.sidefx.com/docs/houdini20.5/nodes/dop/popcurveincompressibleflow.html}{POP Curve Incompressible Flow dynamics node}) in the visual effects software Houdini 20.5~\cite{Houdini} (\cref{fig:teaser}).

\section{Method of Regularized Stokeslets}

\begin{figure}[b]
    \centering
    \includegraphics[width=0.33\linewidth, trim={4.5cm 10cm 3.5cm 10cm},clip]{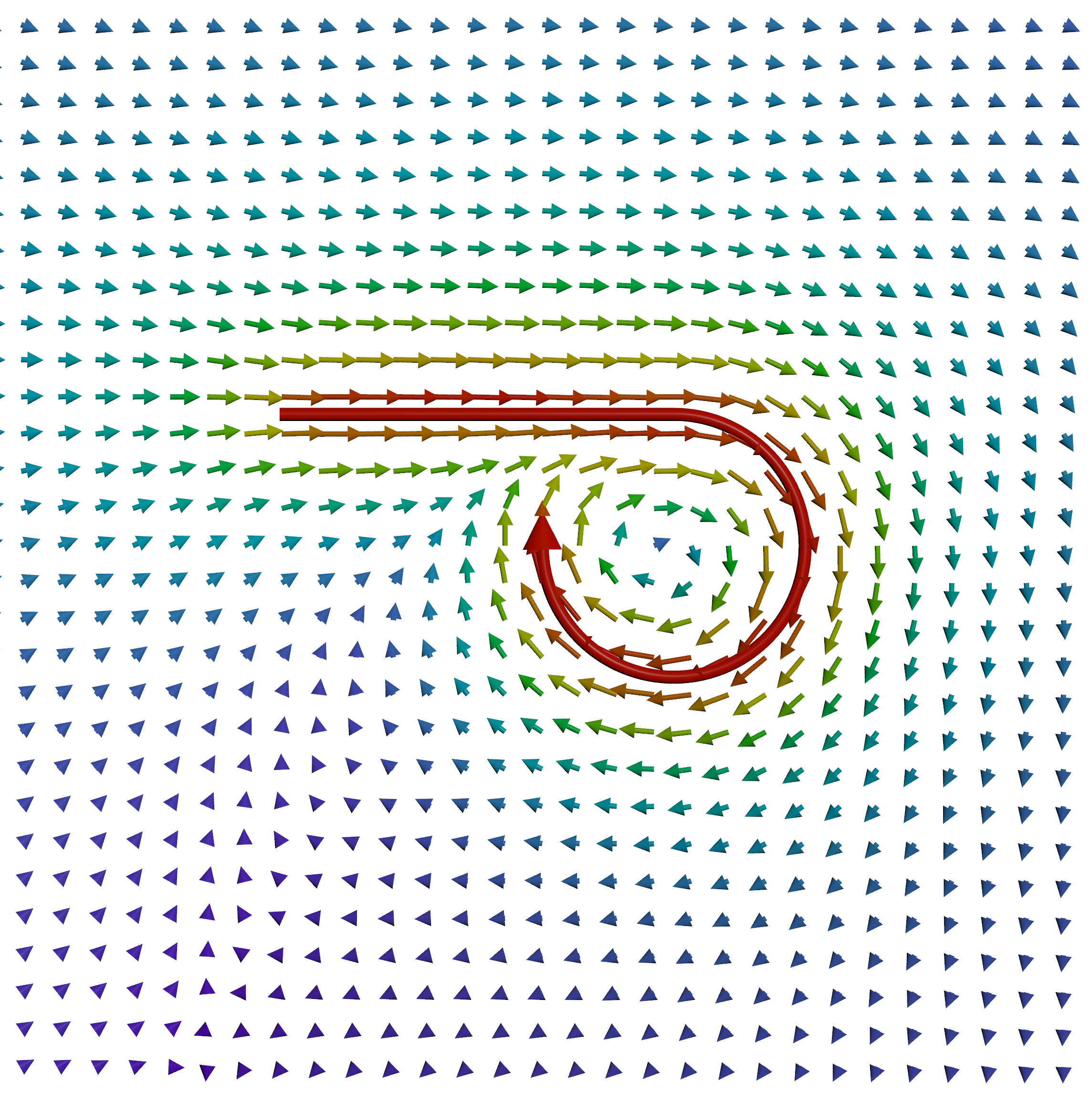}%
    \includegraphics[width=0.33\linewidth, trim={4.5cm 10cm 3.5cm 10cm},clip]{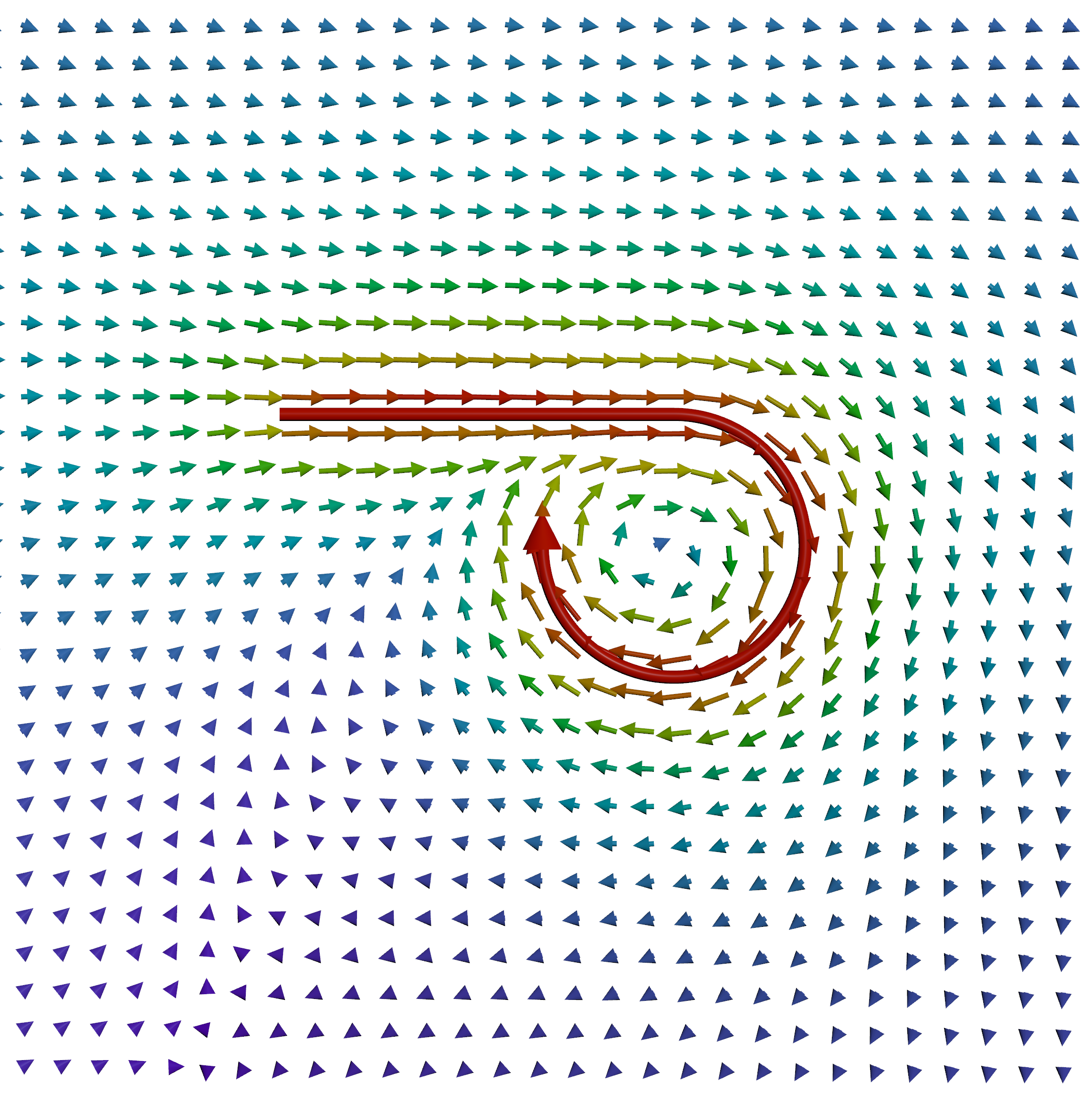}%
    \includegraphics[width=0.33\linewidth, trim={4.5cm 10cm 3.5cm 10cm},clip]{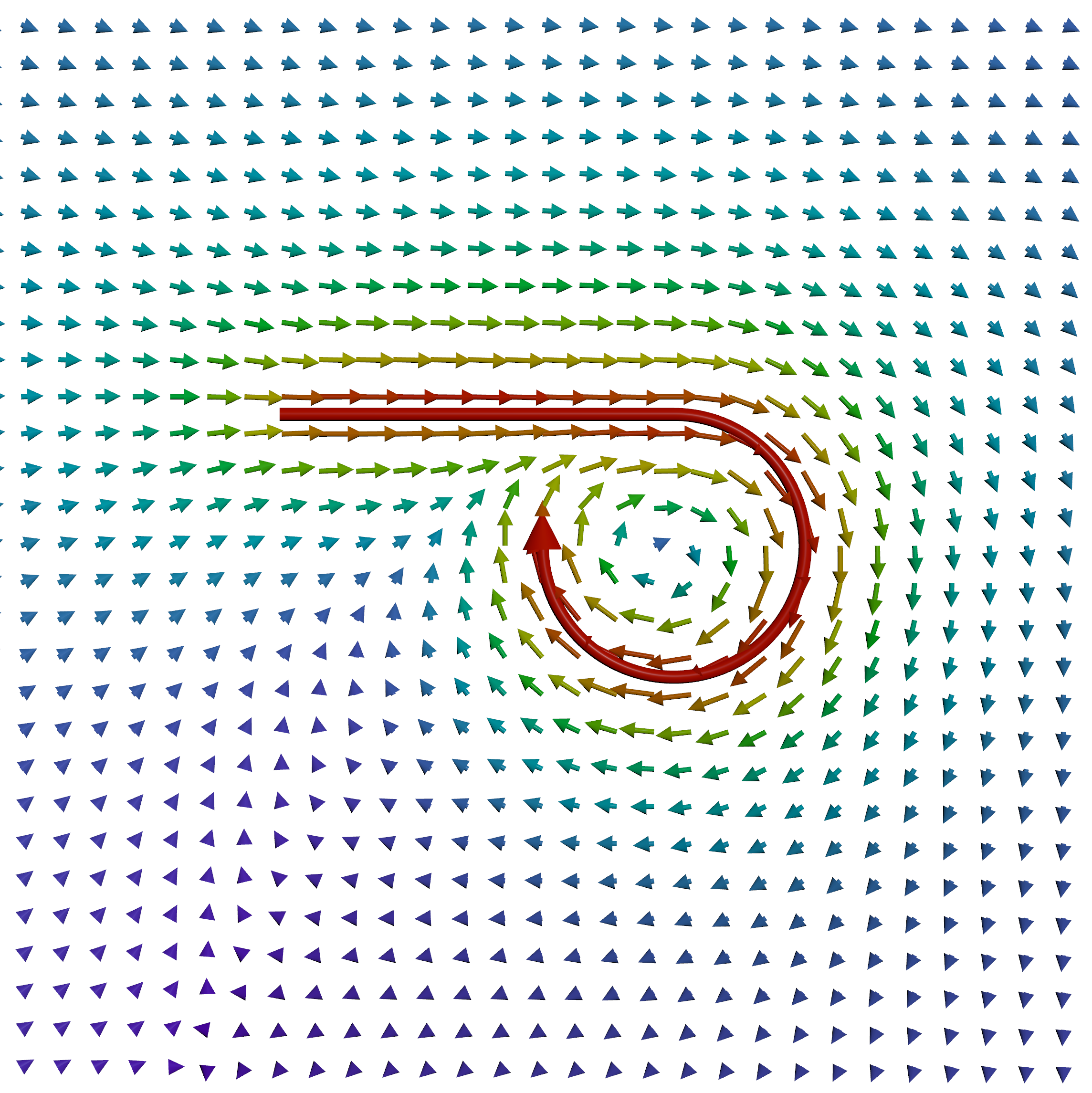}
    \includegraphics[width=0.33\linewidth, trim={4.5cm 10cm 3.5cm 10cm},clip]{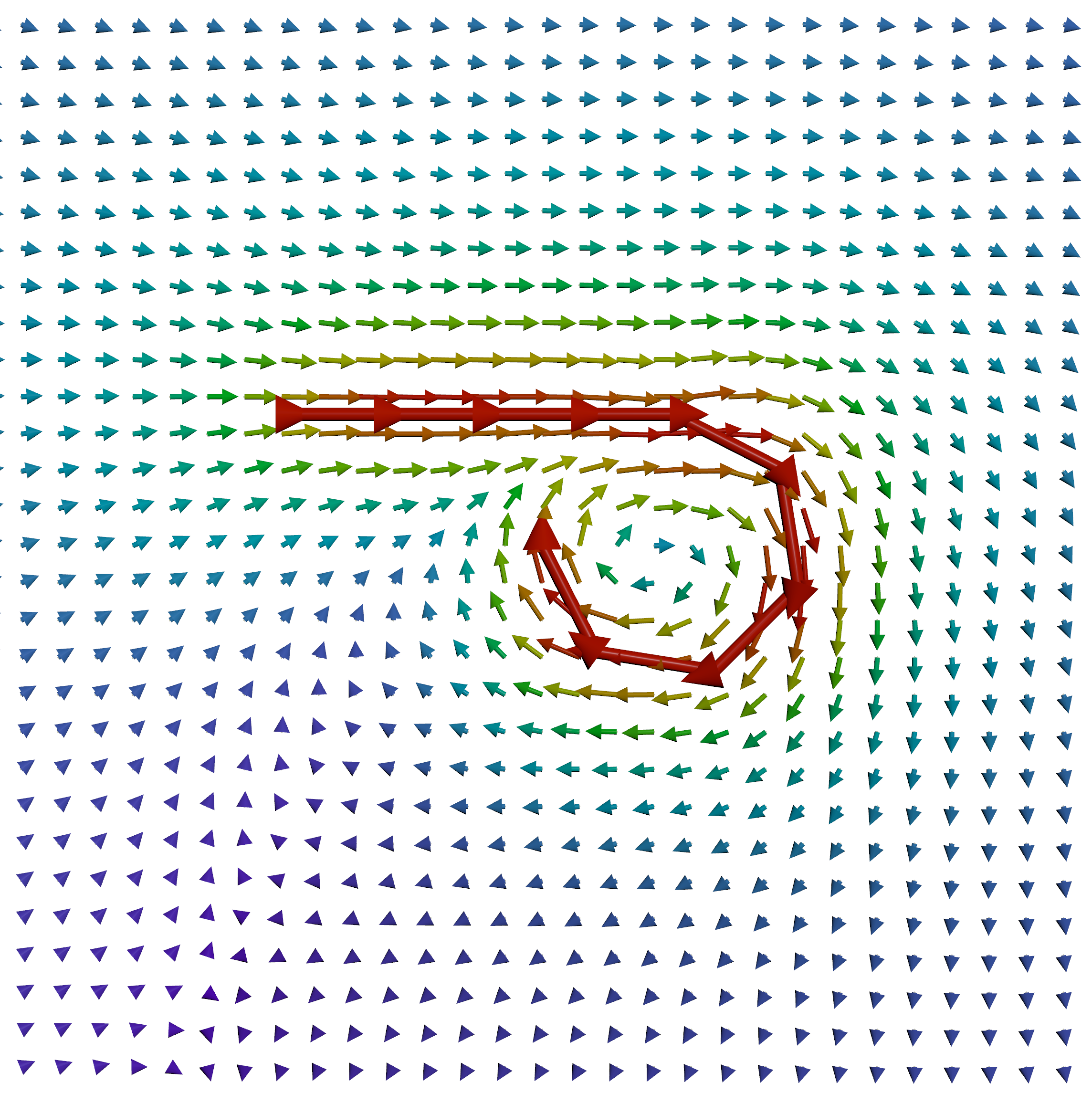}%
    \includegraphics[width=0.33\linewidth, trim={4.5cm 10cm 3.5cm 10cm},clip]{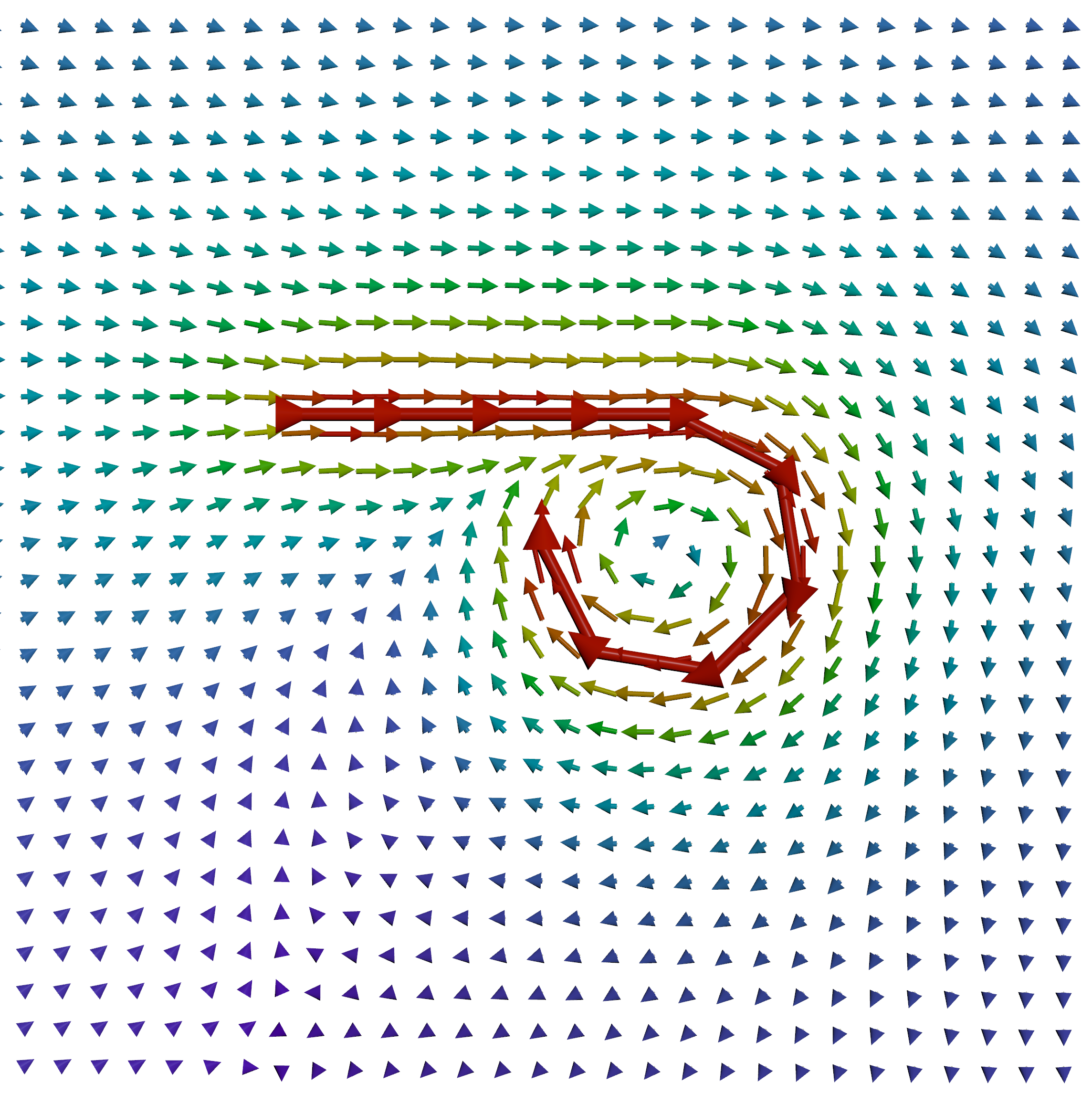}%
    \includegraphics[width=0.33\linewidth, trim={4.5cm 10cm 3.5cm 10cm},clip]{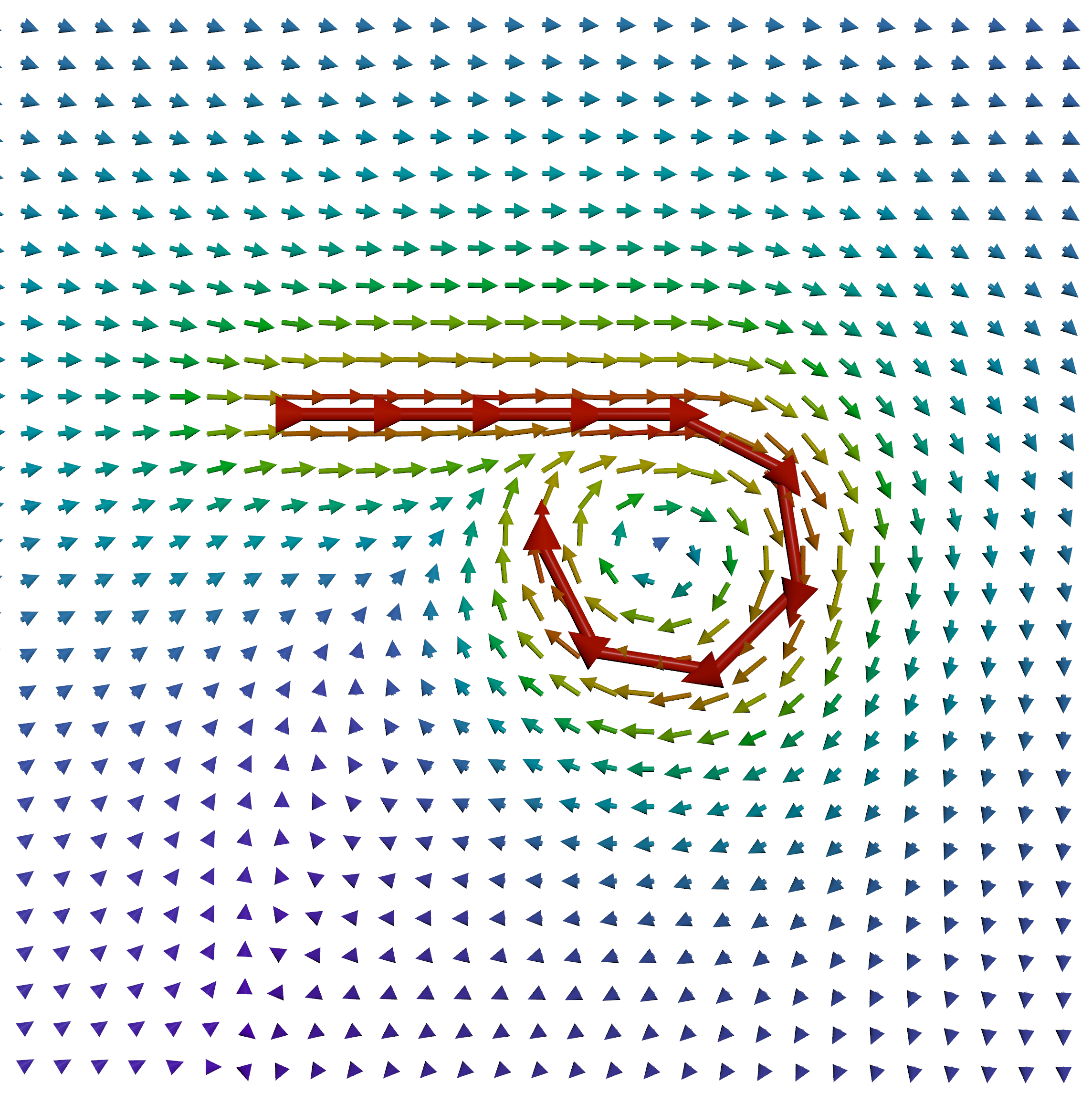}
    \includegraphics[width=0.33\linewidth, trim={4.5cm 10cm 3.5cm 10cm},clip]{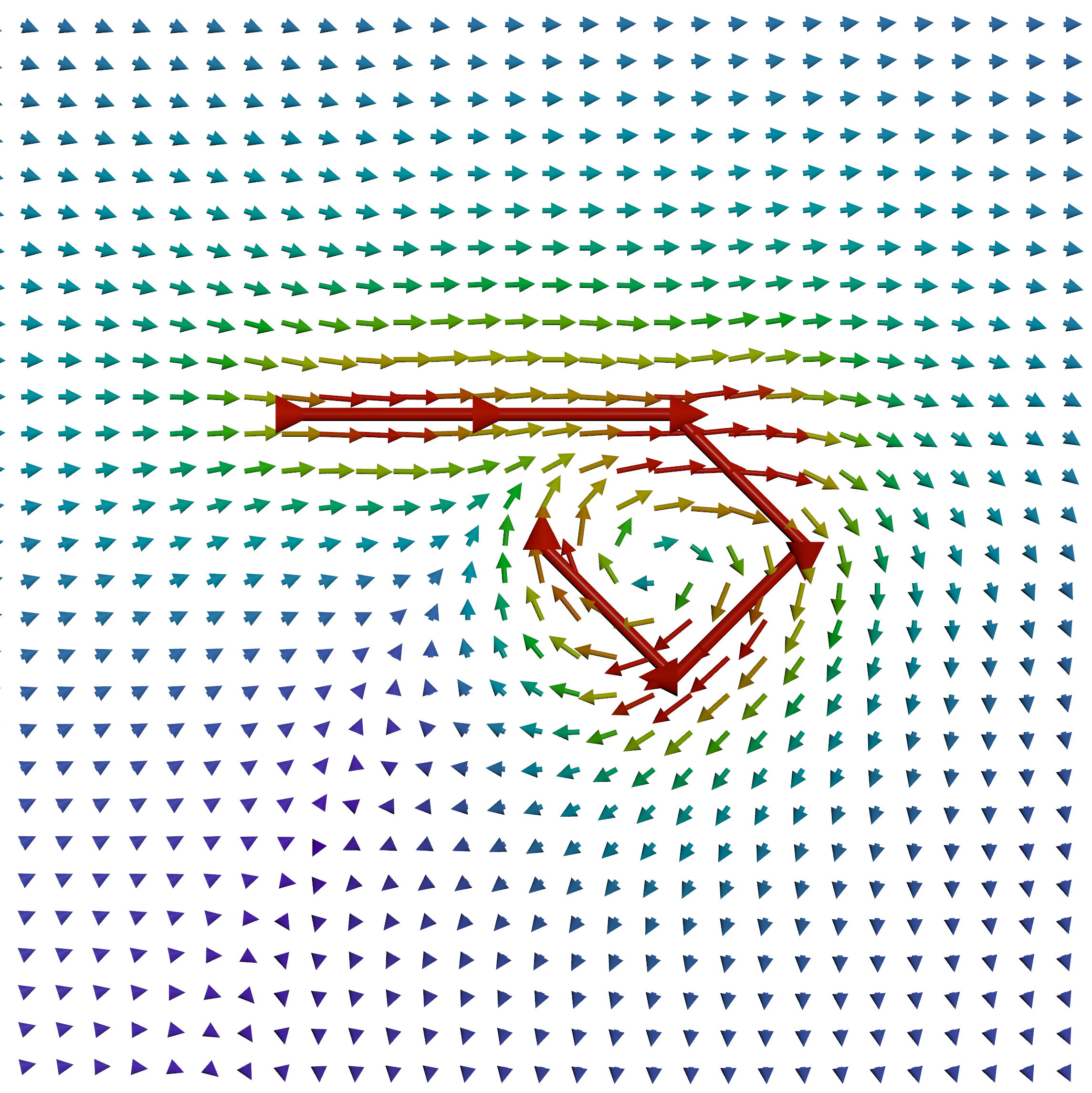}%
    \includegraphics[width=0.33\linewidth, trim={4.5cm 10cm 3.5cm 10cm},clip]{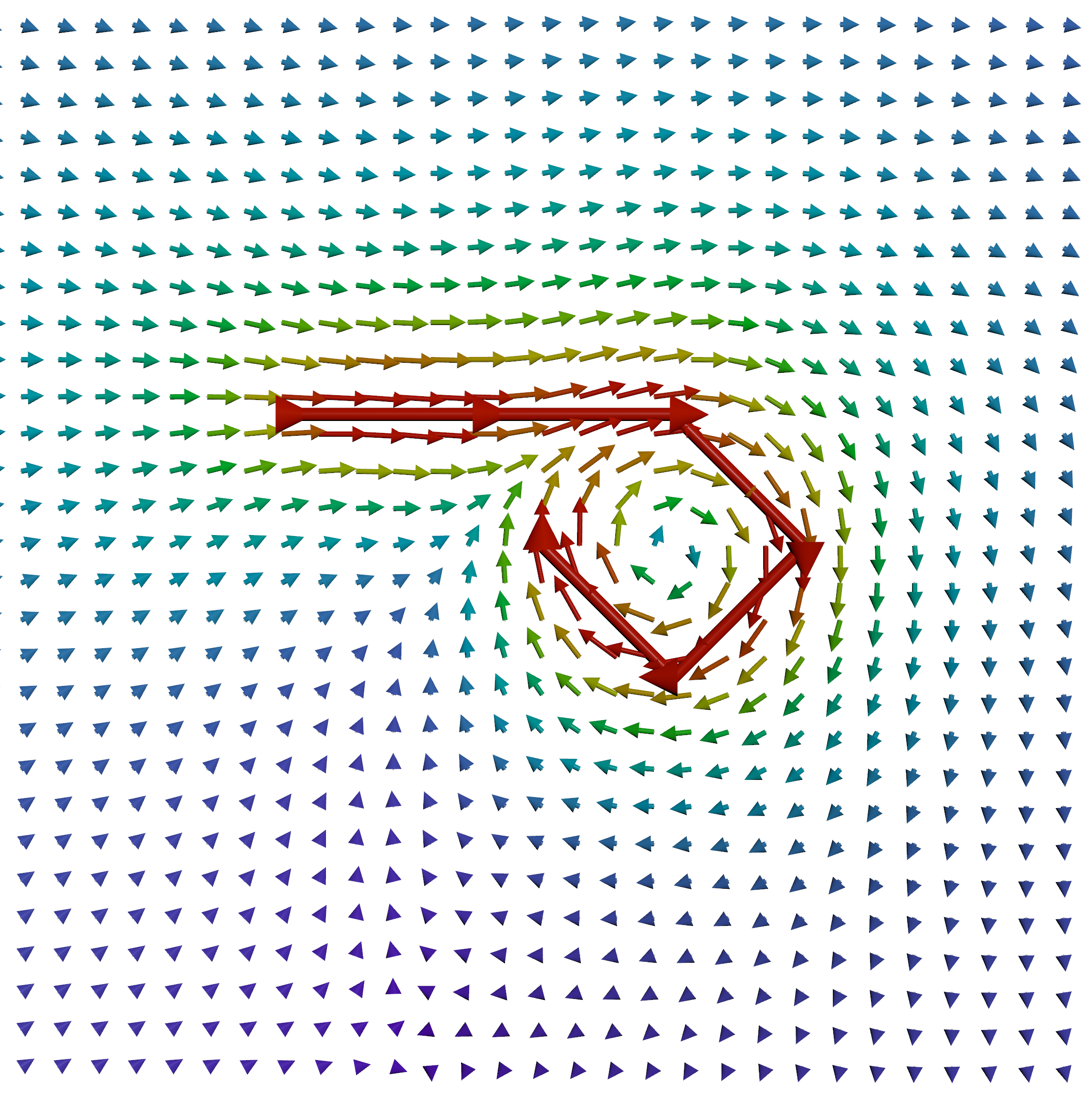}%
    \includegraphics[width=0.33\linewidth, trim={4.5cm 10cm 3.5cm 10cm},clip]{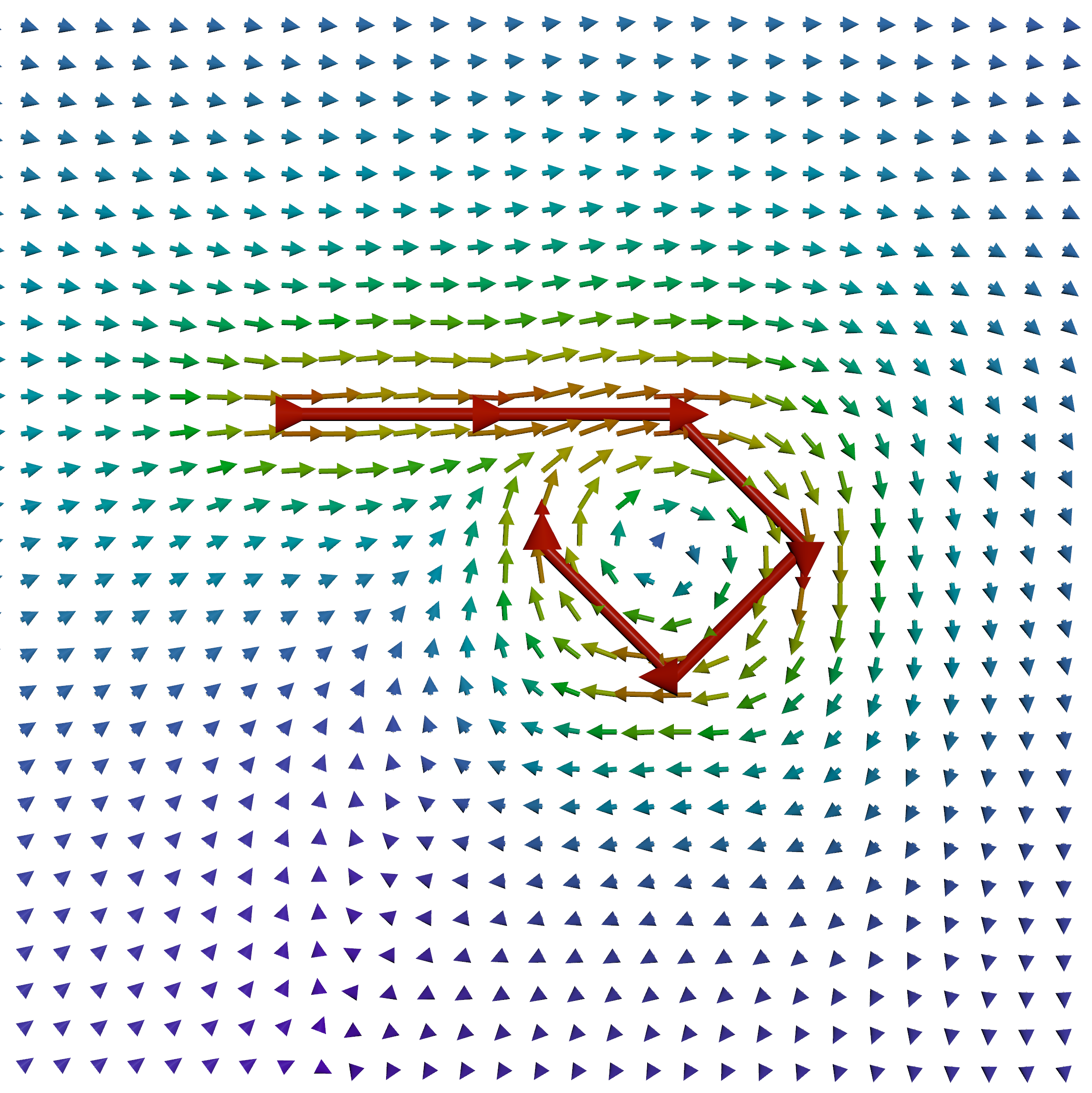}
    \begin{minipage}[b]{0.33\linewidth}\centering\small\textsf{Galerkin (ours)}\end{minipage}%
    \begin{minipage}[b]{0.33\linewidth}\centering\small\textsf{Collocation}\end{minipage}%
    \begin{minipage}[b]{0.33\linewidth}\centering\small\textsf{Point-based}\end{minipage}
    \caption{We compare the velocity field produced by Galerkin (left), collocation (middle), and point-based (right) discretizations with curves having different sampling rates. From the top, we have 512, 11, and 7 vertices along the control curve. For a dense set of points, all three methods perform equally well, but for a sparse set of points, the velocity field from the Galerkin method best follows the control curve. In the bottom row, we can observe that the only velocity we get with Galerkin discretization can respect the horizontal velocity along the long horizontal edge of the input control curve.}
    \label{fig:galerkin}
\end{figure}

The Stokes equations that MRS~\cite{Cortez2001Stokeslets, Cortez2005Stokeslets3D} uses to model the effect of highly viscous and incompressible flow due to a body force term $\mathbf{b}$ are given as
\begin{equation}\label{eq:Stokes}
\mu\nabla^2\vel -   \nabla \pres  + \bodyf = 0\quad\text{and}\quad
    \nabla \cdot\vel = 0,
\end{equation}
where $\vel$ is velocity, $\pres$ is pressure, and $\mu$ is constant dynamic viscosity. As we are only interested in the resulting velocity field, 
we assume $\mu=1$ without loss of generality. Let us consider the solution to the Stokes equations in an unbounded domain, $\Rtwo$ or $\Rthree$.
Suppose that we have a spatially concentrated body force $\bodyf(\vecx) = \delta(\vecx - \vecy)\mathbf{f}$ for \cref{eq:Stokes}. Solutions to PDEs under such concentrated source terms are known as \emph{fundamental solutions}, and for the Stokes equations,
the velocity fundamental solution is called the \emph{Stokeslet}:
$
\stokeslet_{3D}(\vecx, \vecy)  =\frac{1}{8\pi\mu} \left\{ \frac{1}{r}\identity + \frac{1}{r^3}\vecr\vecr\transpose\right\}
$ 
where $\vecr = \vecy - \vecx$, $r = \lVert \vecr\rVert_2$. The resulting velocity field is given as
$\vel(\vecx) = \stokeslet(\vecx, \vecy)\force$.
The Stokeslet is singular at the source location (i.e., unbounded at $\vecx = \vecy$), and using it directly during computation is prone to numerical issues. Instead of this regular Stokeslet, we consider solutions to \cref{eq:Stokes} under a concentrated yet regularized load $\mathbf{b}(\vecx) = \delta\regularized(\vecx - \vecy)\mathbf{f}$. In 3D, we follow the specific choice of regularization $\delta\regularized$ by \citet{Cortez2005Stokeslets3D},
which yields the \emph{regularized Stokeslet} 
\begin{equation}\label{eq:regularized_stokeslet}
    \stokeslet\regularized_{3D}(\vecx, \vecy) = \frac{1}{8\pi\mu} \left\{\frac{r^2+2\epsilon^2}{\re^3}\identity + \frac{1}{\re^3}\vecr\vecr\transpose\right\}
\end{equation}
where $r_\epsilon = \sqrt{r^2 + \epsilon^2}$, $\epsilon$ is a small positive regularization constant, and
$\vel(\vecx) = \stokeslet\regularized(\vecx, \vecy)\force$.
In 2D, we follow the regularization by \citet{DeGoes2017Kelvinlets} to get
\begin{equation}\label{eq:regularized_stokeslet_2d}
    \stokeslet\regularized_{2D}(\vecx, \vecy) = \frac{1}{4\pi\mu} \left\{\left(\frac{\epsilon^2}{\re^2} -\ln{\re}  \right)\identity + \frac{1}{\re^2}\vecr\vecr\transpose\right\}.
\end{equation}
As \cref{eq:Stokes} is linear, we can superpose regularized Stokeslets placed at different positions to express more complex flows. 
Moreover, we can directly specify the velocities at user-defined control points by constructing a linear system and solving for the forces.
However, the baseline MRS outlined above considers velocity and force degrees of freedom located only at discrete (isolated) point locations. Properly modeling velocity constraints applied \emph{continuously} along curves with this naive approach would require sampling points very densely along the curve to achieve the desired quality (\cref{fig:galerkin}), and that comes with a high computational cost because of the associated huge number of degrees of freedom.

\begin{figure*}
    \centering
    \includegraphics[width=0.195\linewidth]{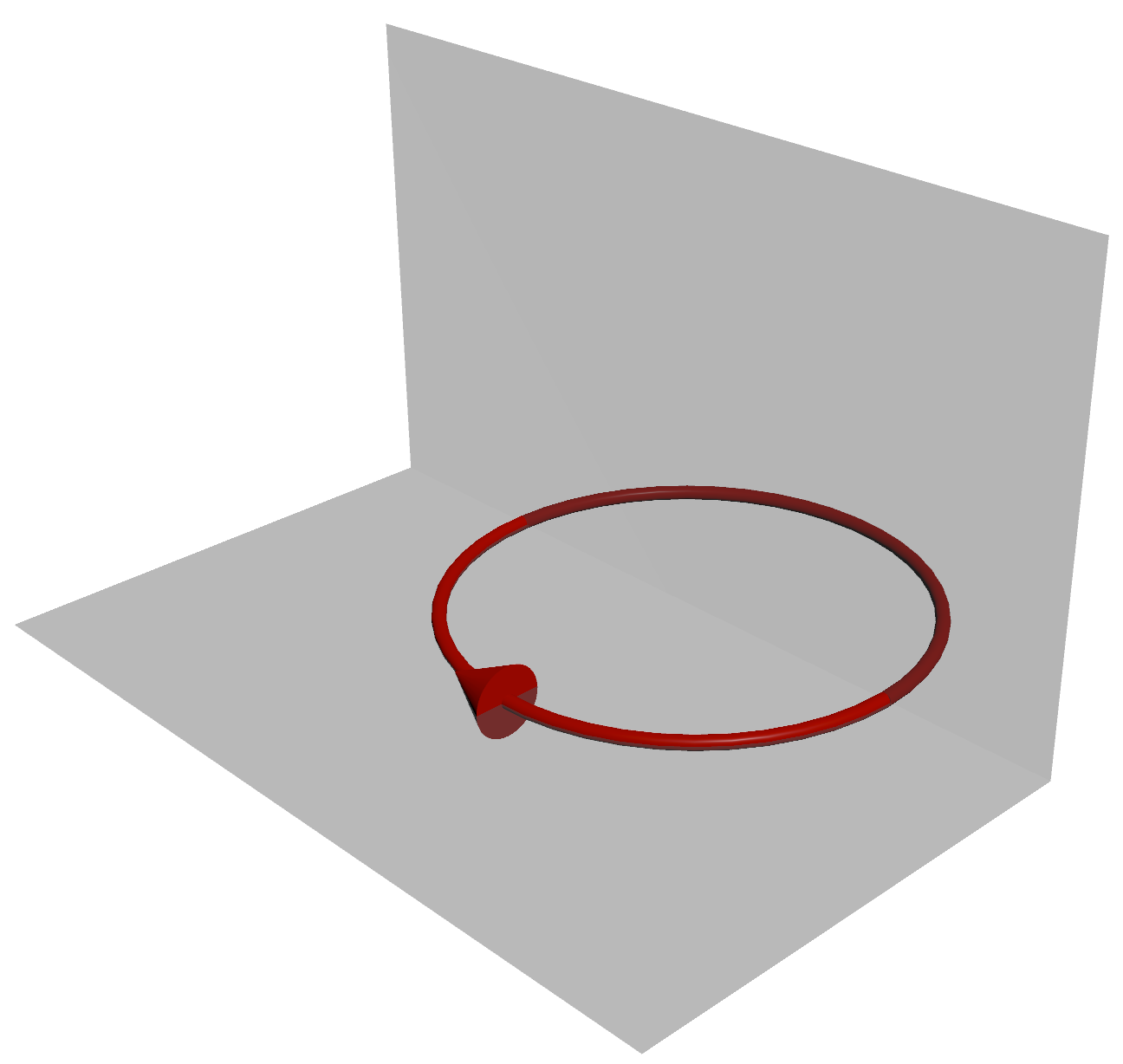}%
    \includegraphics[width=0.195\linewidth]{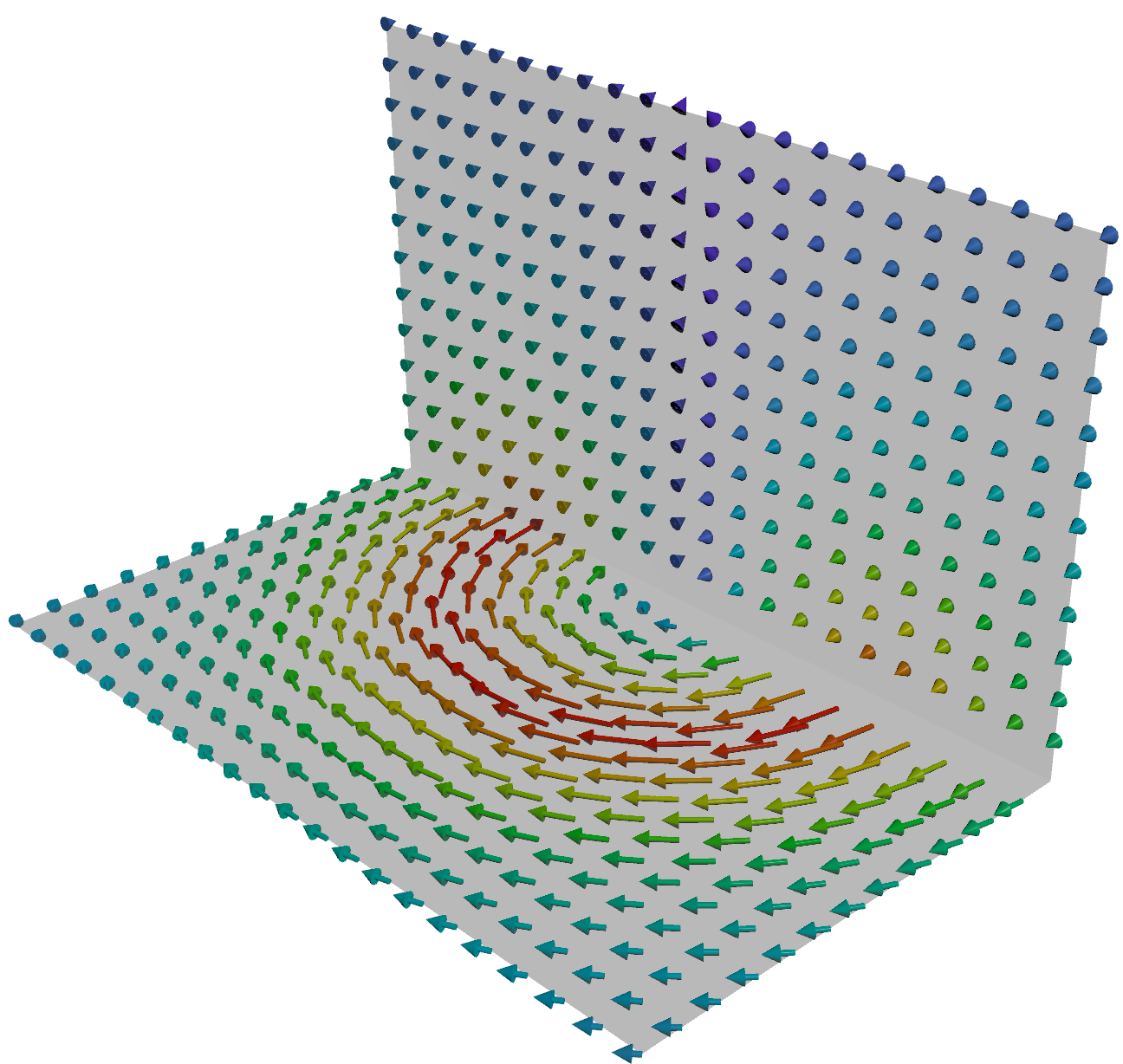}%
    \includegraphics[width=0.195\linewidth]{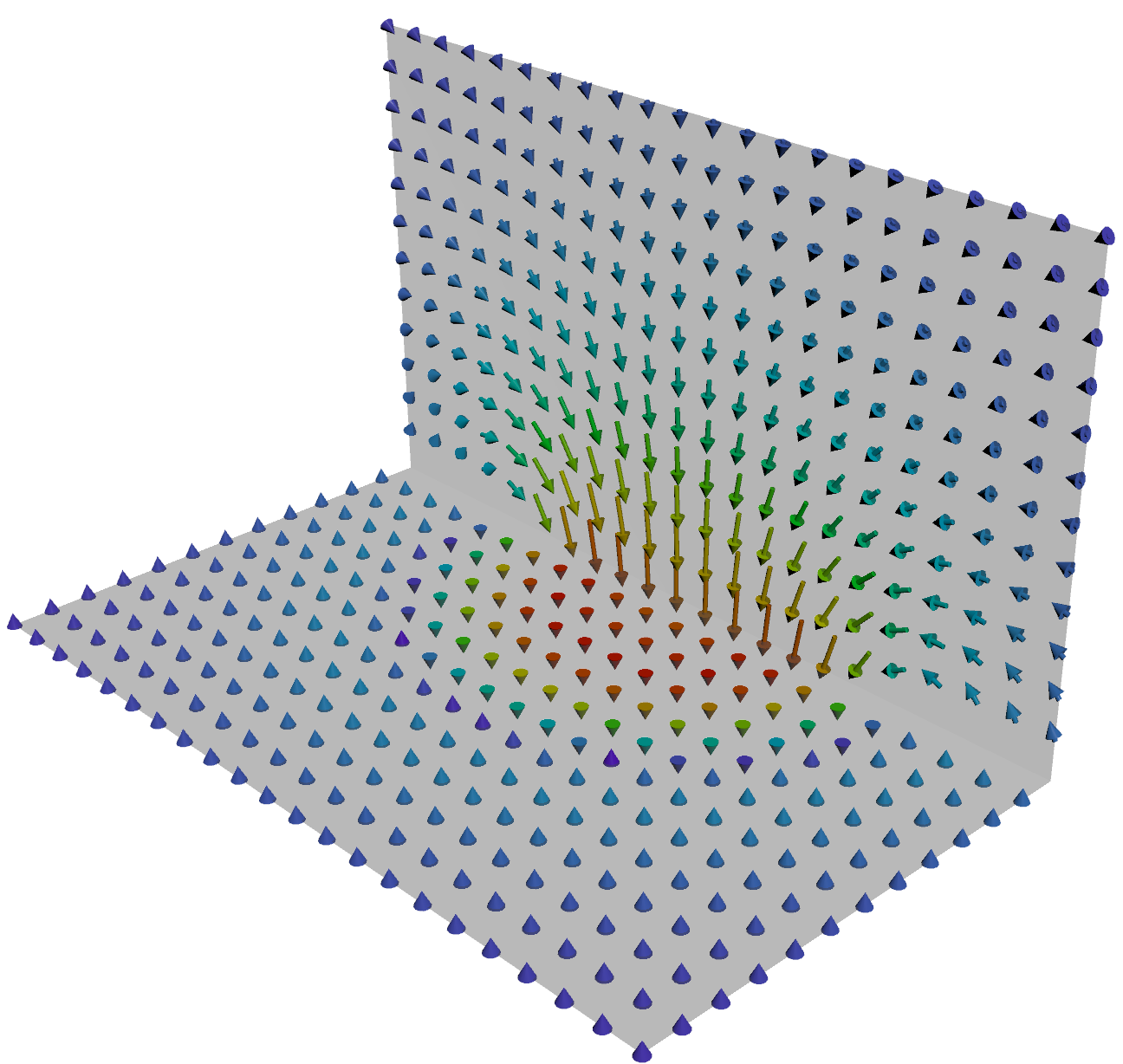}%
    \includegraphics[width=0.195\linewidth]{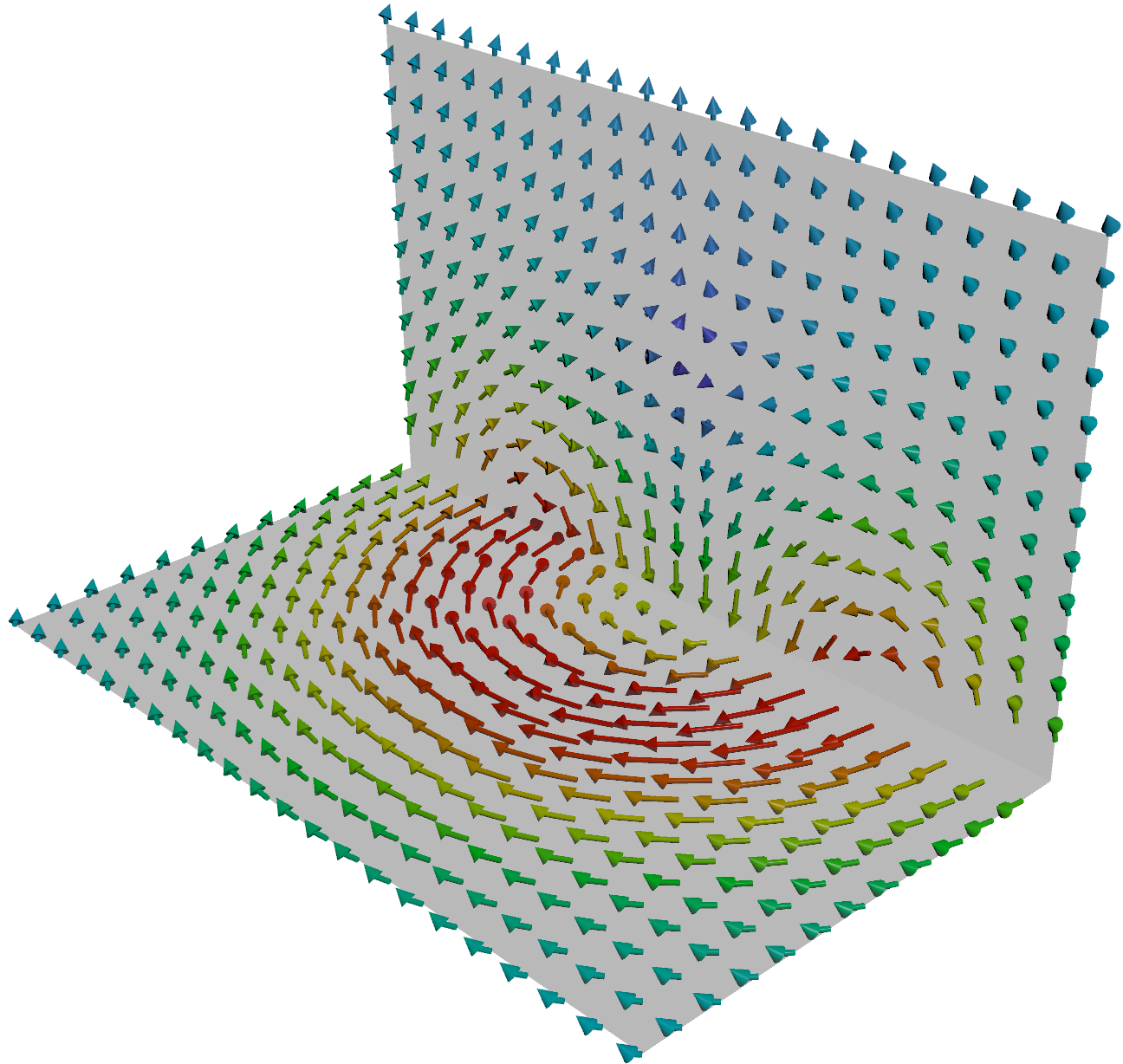}%
    \includegraphics[width=0.195\linewidth]{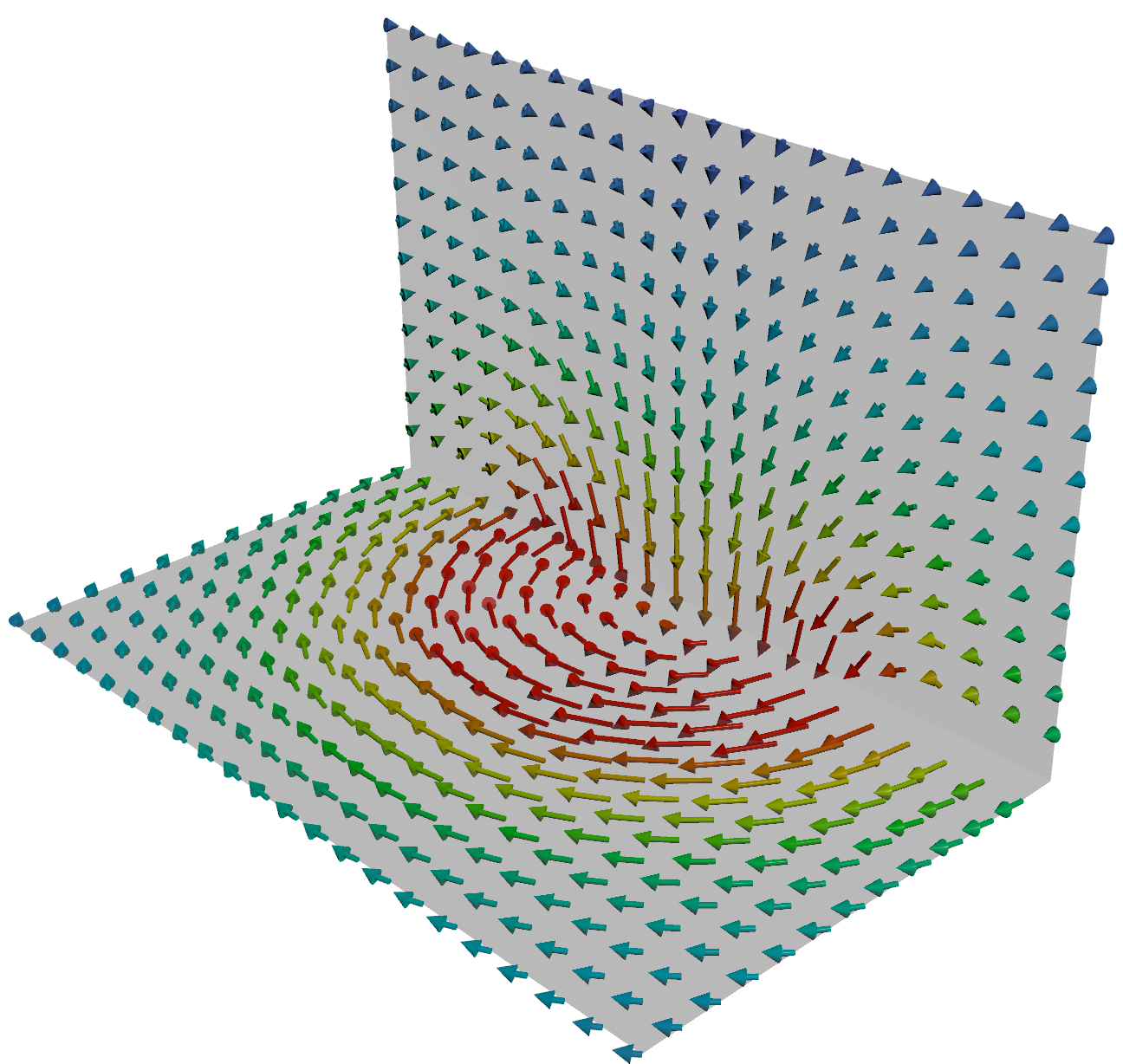}
    \begin{minipage}[b]{0.195\linewidth}\centering\small\textsf{(a) Control Curve}\end{minipage}%
    \begin{minipage}[b]{0.195\linewidth}\centering\small\textsf{(b) Velocity}\end{minipage}%
    \begin{minipage}[b]{0.195\linewidth}\centering\small\textsf{(c) Angular Velocity}\end{minipage}%
    \begin{minipage}[b]{0.195\linewidth}\centering\small\textsf{(d) Coupled}\end{minipage}%
    \begin{minipage}[b]{0.195\linewidth}\centering\small\textsf{(e) Decoupled}\end{minipage}%
    \caption{We specify a velocity of constant magnitude, an angular velocity of constant magnitude, or both, along the control curve (a). We visualize the resulting velocity field on the vertical and horizontal planes by projecting the colored vectors, which represent the velocities, to the two planes. We can specify the velocity and angular velocity independently (b, c) or at the same time (d, e). When we perform the coupled solve of velocity and angular velocity (d), the result slightly differs from the decoupled one (e). While the coupled solve (d) respects the input constraints better, we found that the result of the decoupled solve (e), which outputs the sum of the velocity due to velocity and angular velocity control, is more intuitive given (b) and (c).\looseness=-1}
    \label{fig:velavel}
\end{figure*}

\section{Constraints along Polylines}\label{sec:polylines}
We choose to represent such curves to specify velocities (control curves) as polylines $\curve$ and linearly interpolate the velocity specified at the polyline vertices to define the velocity along them, i.e., $\force(\vecx) = \Phi(\vecx)\mathsf{f}$, where $\Phi(\vecx)$ is the linear interpolation matrix and $\mathsf{f}$ represents the stacked vertex force stored at vertices along the polylines. Once $\mathsf{f}$ is given, one can compute the velocity at any point $\vecx$ by
\begin{equation}\label{eq:reconst}
    \vel(\vecx)  = \left(\int_\curve \stokeslet\regularized(\vecx, \vecy) \Phi(\vecy) \dsy\right) \mathsf{f}.
\end{equation}
One may be tempted to construct a linear system to find the force by requiring that the velocity residual, computed with this equation, evaluate to zero only at the vertex positions of the control curves. However, such a collocation method still yields artifacts when the vertex sampling along the control curves is not dense enough (see \cref{fig:galerkin}). Instead, we consider weighted error residuals along the control curves:
a Galerkin discretization gives the linear system
\begin{equation}\label{eq:galerkin}
    \left(\int_{\curve} \Phi\transpose(\vecx)\Phi(\vecx) \dsx\right)\mathsf{u}
    = \left(\iint_{\curve\times\curve} \Phi\transpose(\vecx) \stokeslet\regularized(\vecx, \vecy) \Phi(\vecy) \dsy\dsx\right) \mathsf{f},
\end{equation}
where $\mathsf{u}$ represents the stacked velocity values at vertices of the polylines. Given $\mathsf{u}$, we first solve \cref{eq:galerkin} to find $\mathsf{f}$. 
Once we find  $\mathsf{f}$, we can evaluate \cref{eq:reconst} to find the velocity at any point in space.

While the regularization parameter $\epsilon$ is typically a constant in MRS, we found that defining it as a function $\epsilon(\vecy)$ over the control curves adds further flexibility in defining velocity fields. Effectively, this varying $\epsilon$ serves as the influence distance.

Note that the related work by \citet{DeGoes2017Kelvinlets} considers the generalization of MRS to elasticity problems and shows several extensions, but they did not consider the constraints on control curves. Thus, our contribution is complementary to theirs.

\section{Angular Velocity Control}
In 3D, it can be desirable to have control over the angular velocity in addition to or instead of the (linear) velocity (\cref{fig:velavel}). We adapt the twist control for elastic displacement control introduced by \citet{DeGoes2017Kelvinlets} to our context. 
While in theory we can constrain the angular velocity relative to any direction vector, we focus on constraining it along the curve's tangent direction.

The velocity field incurred due to a regularized point-concentrated \emph{torque} with magnitude $\torque$ in the tangential direction of the curve at point $\vecy$, $\vect_\vecy$, can be expressed as
\begin{equation}\label{eq:torquetovel}
   \vel(\vecx) = \frac{2\re^2+3\epsilon^2}{4\pi \re^5} \vecr\times\vect_\vecy \,\torque.
\end{equation}
We can also obtain the corresponding angular velocity field $\boldsymbol\omega(\vecx)$ by taking the curl of \cref{eq:torquetovel} and multiplying it by $0.5$:
\begin{equation}\label{eq:torquetoavel}
\begin{split}
\boldsymbol\omega(\vecx) = -\frac{1}{8\pi \re^7} \left\{ (10\epsilon^4 - 7\epsilon^2 r^2 - 2r^4) (\vect_\vecx\cdot\vect_\vecy) \right.\\
                 \left. + (21 \epsilon^2 + 6r^2) (\vecr\cdot\vect_\vecx)(\vecr\cdot\vect_\vecy) \right\}\torque
\end{split}
\end{equation}
This expression allows the direct control of angular velocity along the curves, analogous to how we can directly specify the velocity along the curves. We can also derive the angular velocity along due to the regularized point-concentrated force by taking the curl of $\vel(\vecx) = \stokeslet(\vecx, \vecy)\force$ and multiplying it by $0.5$:
\begin{equation}\label{eq:forcetoavel}
    \boldsymbol\omega(\vecx) =  \frac{2\re^2+3\epsilon^2}{16\pi \re^5}  (\vecr\times\vect_\vecx)\transpose\,\force.
\end{equation}
Just like \Cref{sec:polylines}, we use Galerkin discretization to find the force and torque along the curve, such that the resulting velocity and angular velocity conform to those specified along the curve.
We can augment the linear system, which describes the relationship between velocity and force (\cref{eq:galerkin}), with these additional relationships between velocity and torque (\cref{eq:torquetovel}), torque and angular velocity (\cref{eq:torquetoavel}), and force and angular velocity (\cref{eq:forcetoavel}). Once we find the force and torque along the curve, to get the resulting velocity field at any point in the domain, in addition to the effect of force on velocity in \cref{eq:reconst}, we compute the effect of torque on velocity as well. We can similarly compute the angular velocity.

We observed that decoupling the velocity-force and angular velocity-torque solves and adding their effects later may give more intuitive control, so we also offer such an option in our tool.

\section{Computational Complexity}
Consider solving the Galerkin-discretized systems, such as \cref{eq:galerkin}. We can evaluate the left-hand side, which contains only the known quantities (velocity and angular velocity), in $\mathcal{O}(N)$ time and memory, where $N$ is the number of vertices in the input polylines, since we can compute the effect of matrix multiplication directly without explicitly forming the matrix.
For the right-hand side, we evaluate the right-hand-side matrix explicitly, which costs $\mathcal{O}(N^2)$ time and memory. This matrix is dense, and solving the linear system costs $\mathcal{O}(N^3)$ computation. While prior work on related techniques~\cite{
Chen2024} discuss this cost associated with dense matrix operations as a limitation, it is not a significant problem in our case as we only have curves as the integral domains, and the problem size is typically small. For our typical application of authoring a static velocity field, the computation of the unknown quantities (force and torque) occurs only once.  %
Once we find the force (and torque), reconstructing the velocity (and angular velocity) costs $\mathcal{O}(NM)$, where $M$ is the number of evaluation points. While $M$ can be large, $N$ is typically small.

We used a 3-point Gaussian quadrature to evaluate integrals over each line segment. As most of the computation is trivially parallelizable, we implemented most parts of the algorithm on GPU using OpenCL, except for the linear system solve.  

\section{Results}
\begin{figure}[b]
    \centering
    \includegraphics[width=0.33\linewidth]{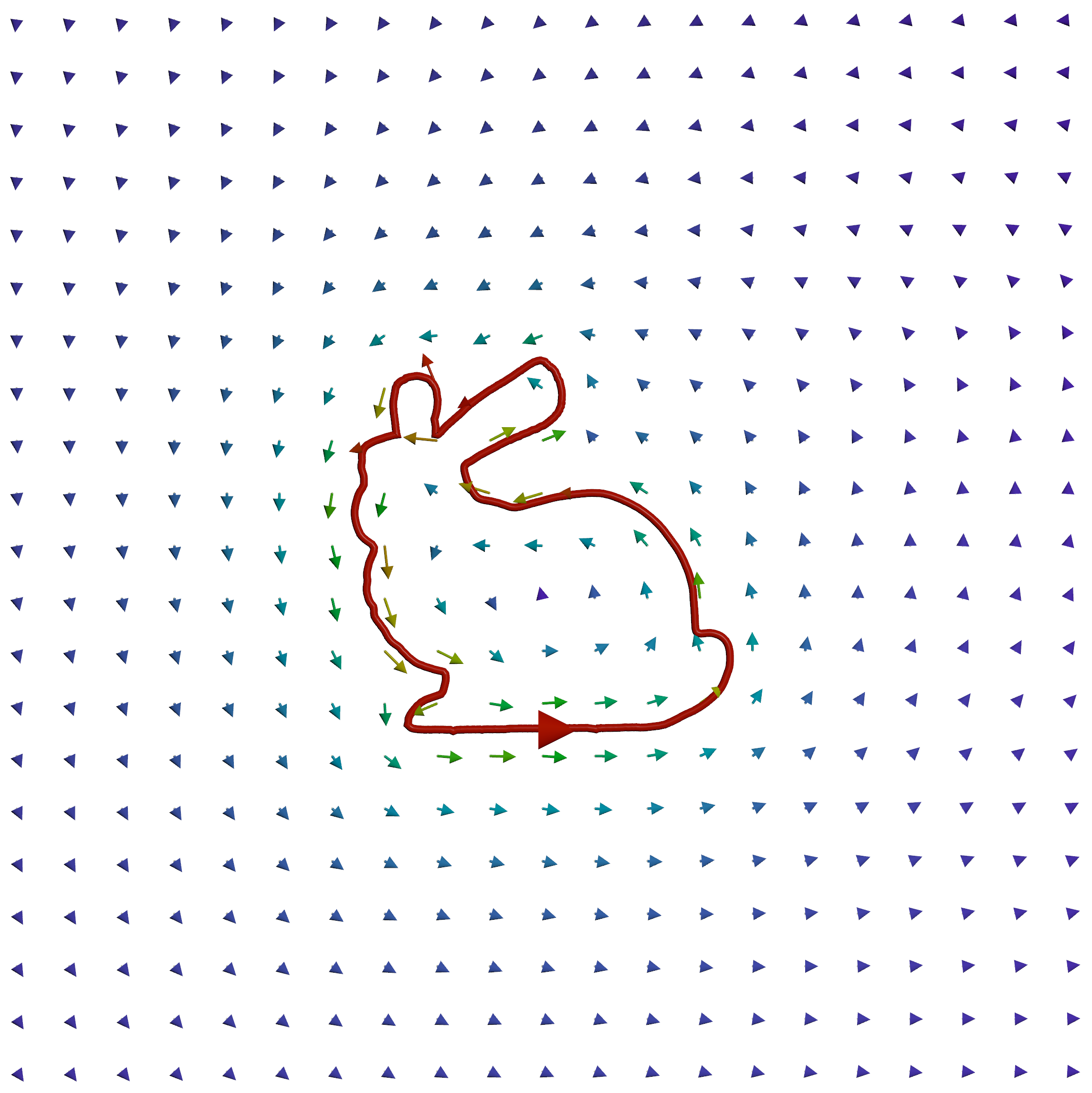}%
    \includegraphics[width=0.33\linewidth]{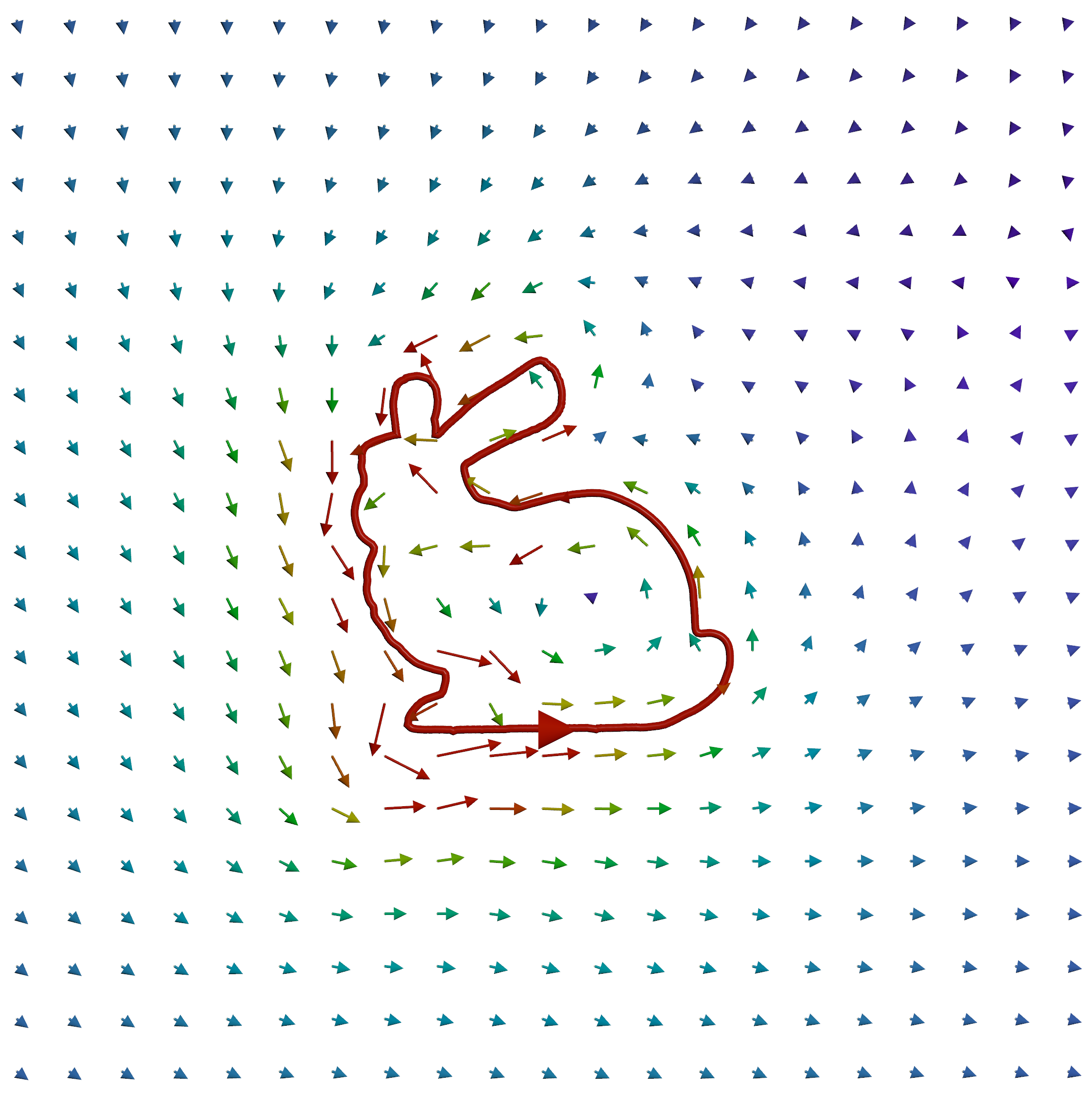}%
    \includegraphics[width=0.33\linewidth]{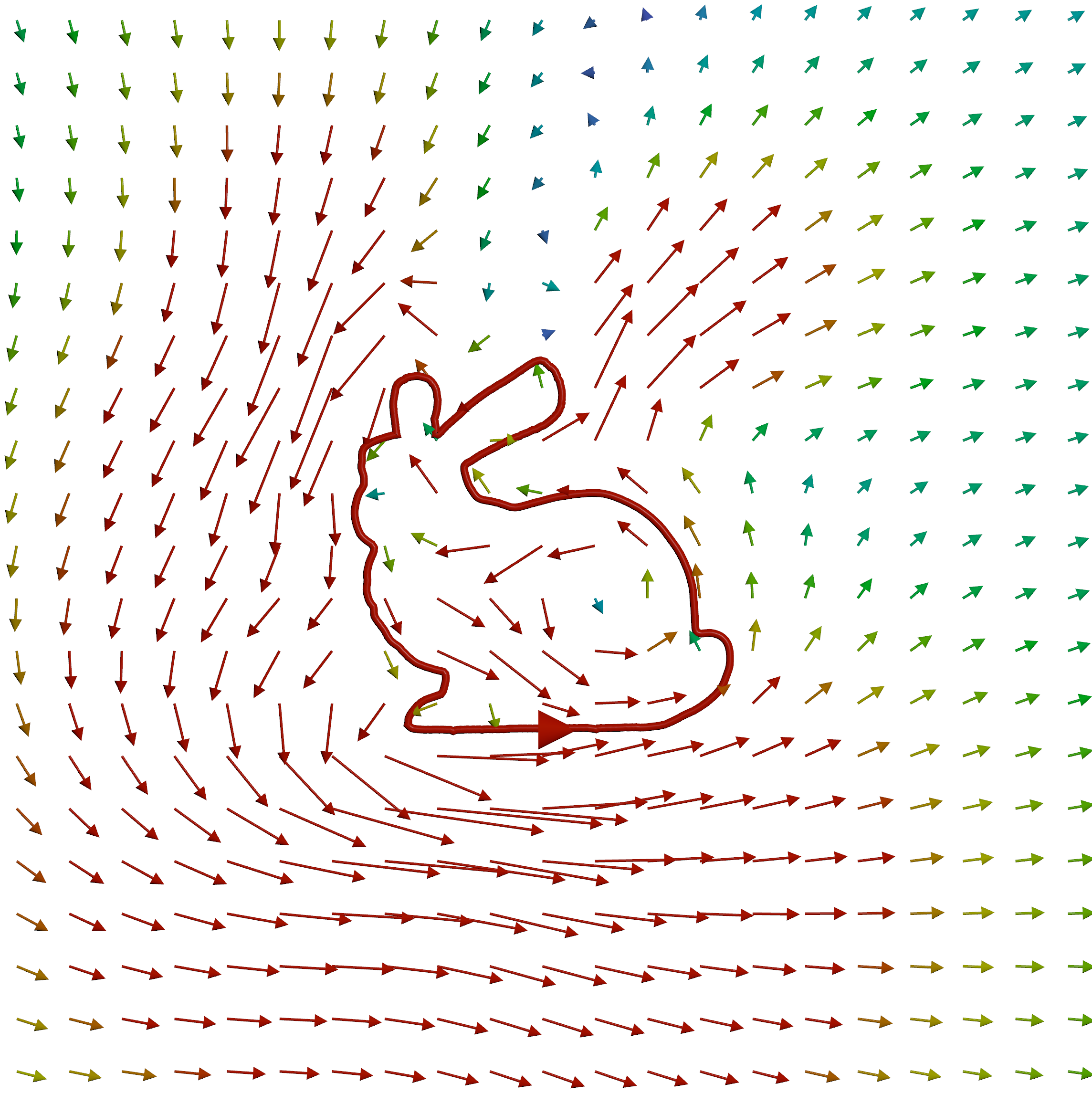}
    \begin{minipage}[b]{0.33\linewidth}\centering\small$\epsilon=0.01$\end{minipage}%
    \begin{minipage}[b]{0.33\linewidth}\centering\small$\epsilon=0.05$\end{minipage}%
    \begin{minipage}[b]{0.33\linewidth}\centering\small$\epsilon=0.10$\end{minipage}%
    \caption{We specify a constant magnitude velocity along the bunny curve with three different constant $\epsilon$ values. Changing the $\epsilon$ value effectively changes the influence distance.}
    \label{fig:radius}
\end{figure}

\begin{figure}
    \centering
    \includegraphics[width=0.33\linewidth]{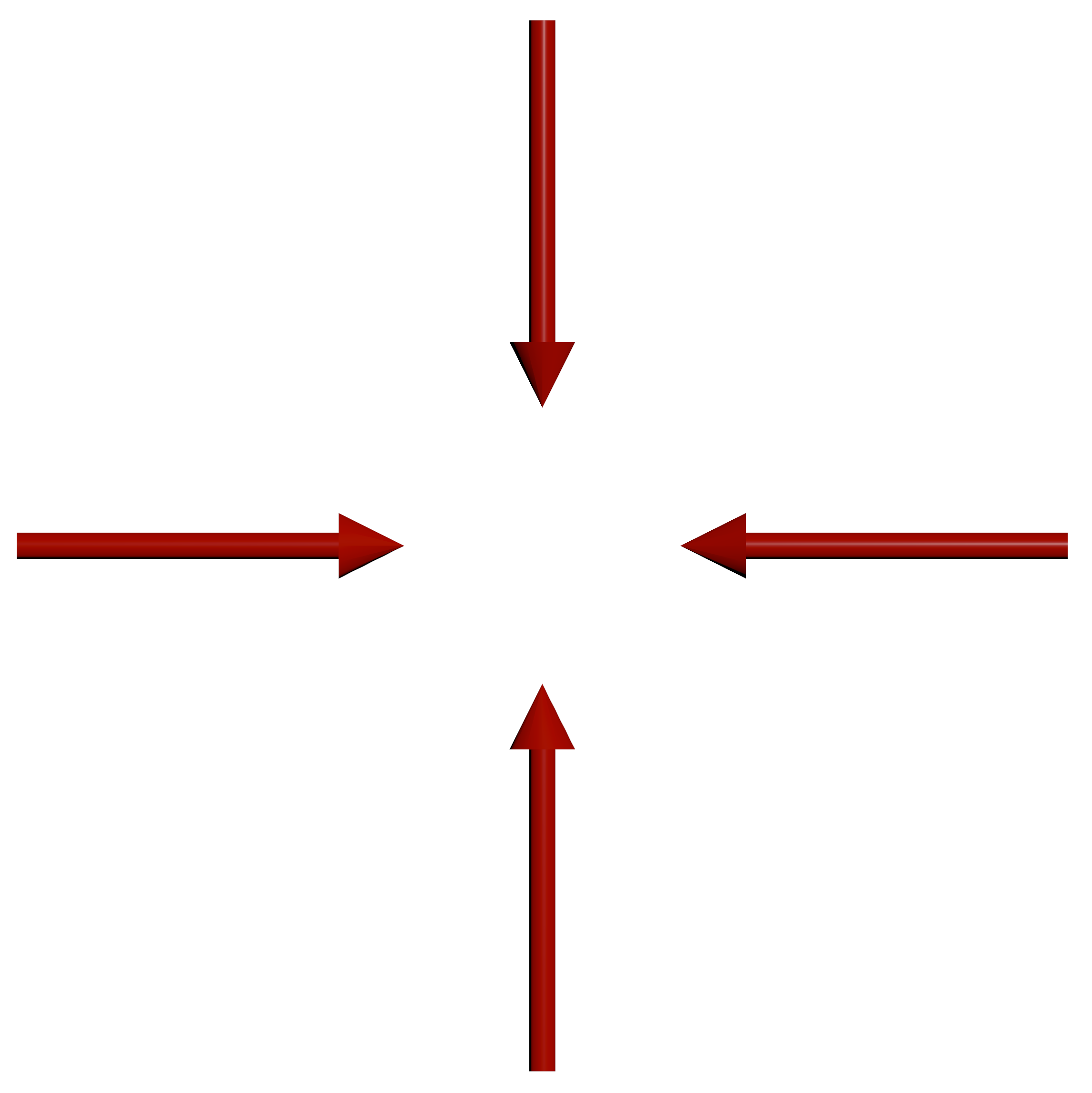}%
    \includegraphics[width=0.33\linewidth]{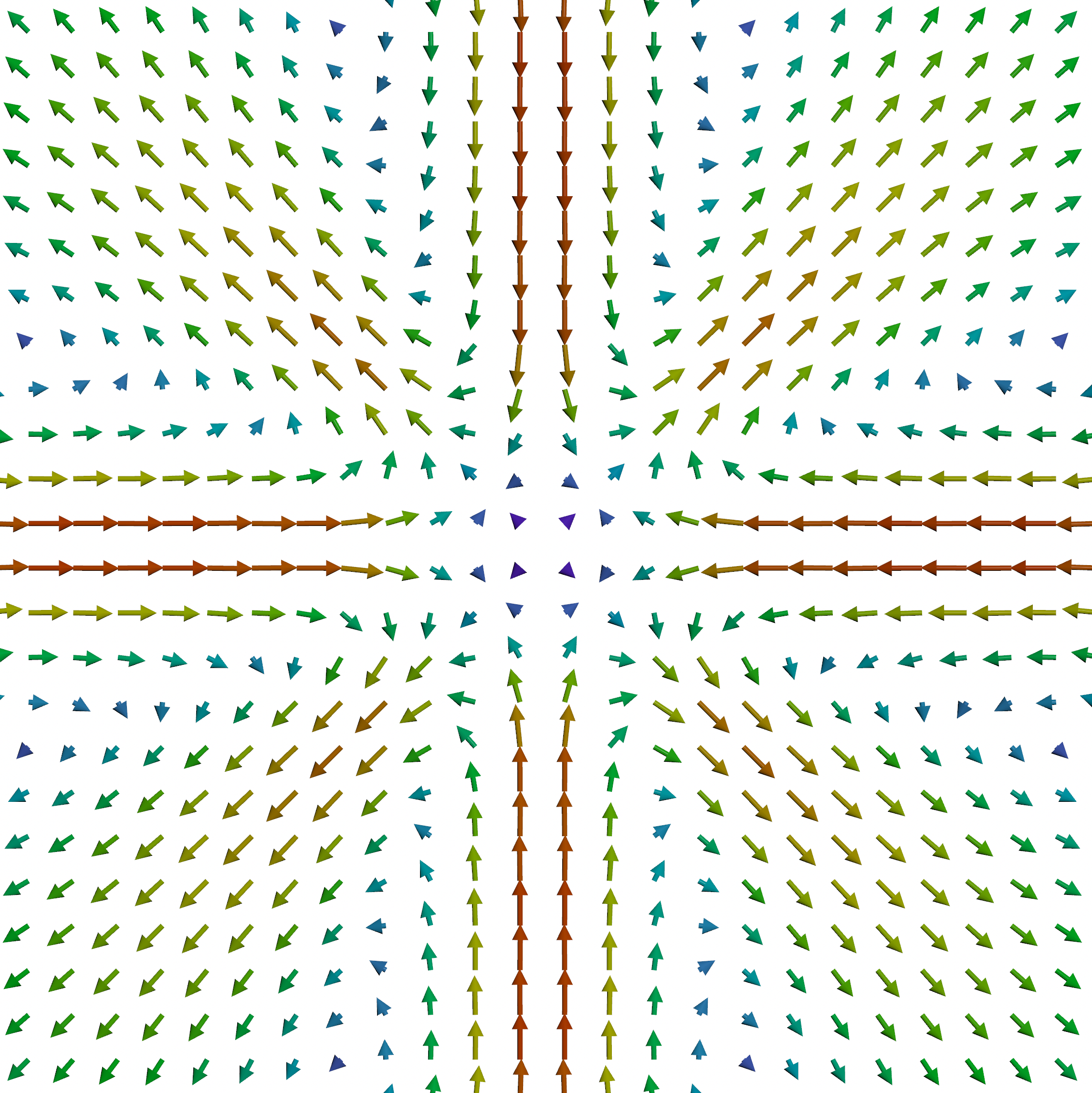}%
    \includegraphics[width=0.33\linewidth]{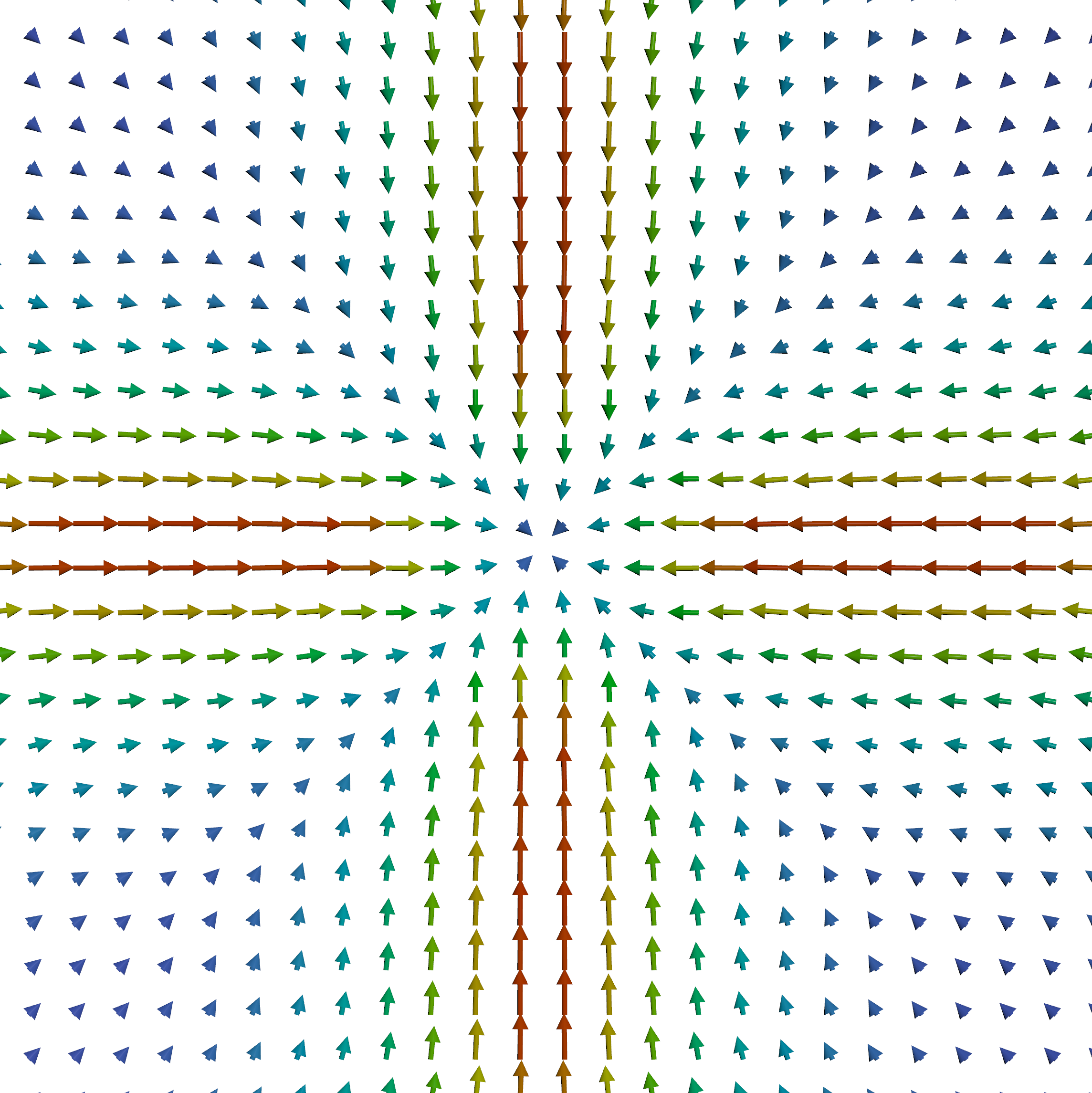}
    \caption{Given four control curves with constant magnitude velocity constraints pointing toward the center (left), we use our 2D version (middle) and 3D version (right) of our method to get an incompressible velocity field, respectively. With the 3D version, the field is not incompressible in the 2D slice.}
    \label{fig:2d}
\end{figure}

Our method can specify the velocity and angular velocity independently or at the same time via a coupled solve or a decoupled solve (\cref{fig:velavel}). 
We can change the regularization parameter $\epsilon$ to control the velocity and angular velocity influence distance from the control curves (\cref{fig:radius}).
\Cref{fig:2d} demonstrates velocity control in 2D space. Applying the 3D version of our method in 2D still gives a valid incompressible field in 3D, but it may not necessarily give a 2D incompressible field. Thus, the 2D version is preferred for 2D scenes.
The (pre)computation of forces in \cref{fig:teaser} took about 0.06 seconds, and the evaluation of velocities at the leaf positions took about 0.003 seconds per frame on a MacBook Pro with M1 Pro. There are 290 vertices along the control curve and 581 velocity evaluation points in this scene.
We provide the Houdini project file used to generate the results in the paper as supplemental material.\looseness=-1

\section{Conclusion and Discussion}
We developed a method to design an incompressible velocity field based on polylines with velocity and optionally angular velocity specified along them. Combining the method of fundamental solutions with Galerkin discretization allowed for intuitive control of the velocity field while limiting the degrees of freedom so that the dense matrix operation cost does not grow significantly.

Our method could naturally be extended for velocity specified over surfaces such as triangle meshes to enable no-slip boundary behavior over solid obstacles in the scene. One can also consider adding normal velocity constraints for free-slip boundaries by augmenting our method with the fundamental solutions for the Laplace equation, following a potential flow assumption. Our early prototype of such extensions found that the accuracy of the result depends heavily on the mesh resolution. Moreover, the $1D\times1D$ integral domain in \cref{eq:galerkin} would become $2D\times2D$ for the Galerkin discretization of such extensions, which would significantly increase the computational cost. 
Further work is needed to make such extensions more practical. 

With our method, the resulting velocity field has a fixed falloff as $r\rightarrow \infty$.
We could adapt the regularized fundamental solutions with faster falloff \cite{DeGoes2019SharpKelvinlets} if desired. For a slower falloff, we could achieve this by setting zero traction boundary conditions on a large sphere that fully contains the computational domain; however, this method comes with the discretization resolution and computational cost problems mentioned in the previous paragraph.

To support velocity or traction constraints over surfaces, using a fine surface discretization with acceleration techniques such as the fast multipole method~\cite{Greengard1997FMM} with a carefully designed preconditioner~\cite{Chen2024} to handle these cases would be an interesting future direction.

\begin{acks}
The majority of this project has been completed while the first author was employed by Side Effects Software Inc.
This research was partially funded by NSERC Discovery Grants (RGPIN-2021-02524 \& RGPIN-2020-03918), CFI-JELF (Grant 40132), and a grant from Autodesk. 
The first author was partially funded by the David R. Cheriton Graduate Scholarship. 
\end{acks}
 
\bibliographystyle{ACM-Reference-Format}
\bibliography{main}

\end{document}